\documentclass[prd,aps,showpacs,floatfix,nofootinbib]{revtex4}
\usepackage{epsfig}
\usepackage{placeins}
\usepackage{amssymb}
\usepackage{color}
\usepackage{float}
\usepackage{hyperref}
\usepackage{yfonts}
\usepackage{amsmath}
\usepackage{MnSymbol}
\usepackage{amssymb}
\usepackage{amsfonts}
\usepackage{eufrak}
\usepackage{bm}
\usepackage{phoenician}
\usepackage{graphicx, calc}
\usepackage{xcolor}
\usepackage{times}
\usepackage{natbib}

\usepackage{soul}
\usepackage{ulem}

\newcommand{\be}{\begin{equation}}
\newcommand{\ee}{\end{equation}}

\newcommand{\bea}{\begin{eqnarray}}
\newcommand{\eea}{\end{eqnarray}}
\newcommand{\bean}{\begin{eqnarray*}}
\newcommand{\eean}{\end{eqnarray*}}

\usepackage{calc}
\newlength{\depthofsumsign}
\makeatletter
\newcommand*{\DivideLengths}[2]{%
  \strip@pt\dimexpr\number\numexpr\number\dimexpr#1\relax*65536/\number\dimexpr#2\relax\relax sp\relax
}
\makeatother

\begin{document}

\title{Disantangling the effects of Doppler velocity and primordial non-Gaussianity in galaxy power spectra}

\author{L. Raul Abramo$^1$, Daniele Bertacca$^2$}
\affiliation{
$^1$ Departamento de F\'{\i}sica Matem\'atica, Instituto de F\'{\i}sica, 
Universidade de S\~{a}o Paulo, Rua do Mat\~{a}o 1371, CEP 05508-090, S\~ao Paulo, Brazil \\
$^2$ Argelander-Institut f\"ur Astronomie, Auf dem H\"ugel 71, D-53121 Bonn, Germany}
 
\date{\today}

\begin{abstract}
We study the detectability of large-scale velocity effects on galaxy clustering, by simulating galaxy surveys and combining the clustering of different types of tracers of large-scale structure.
We employ a set of lognormal mocks that simulate a $20.000$ deg$^2$ near-complete survey up to $z=0.8$, 
in which each galaxy mock traces the spatial distribution of dark matter of that mock with a realistic bias prescription.
We find that the ratios of the monopoles of the power spectra of different types of tracers carry most of the information 
that can be extracted from a multi-tracer analysis.
In particular, we show that with a multi-tracer technique it will be possible to detect velocity effects with $\gtrsim 3 \sigma$.
Finally, we investigate the degeneracy of these effects with the (local) non-Gaussianity parameter $f_{\rm NL}$, 
and how large-scale velocity contributions could be mistaken for the signatures of primordial non-Gaussianity. 
\end{abstract}


\maketitle

\section{Introduction}
Modern cosmology is counting on galaxy surveys to resolve many outstanding questions, from the origin of cosmic acceleration to neutrino masses and inflationary parameters.
In order to answer those questions, high accuracy and precision must be achieved, both in theory and in observations, especially on the largest scales probed by our past light-cone.
At low redshifts, where observations are cleanest, this means not only large distances, but large angles as well~\cite{Bertacca:2012, Raccanelli:2016doppler, Borzyszkowski:2017ayl}.
In this case, there is the need to include corrections to the modeling of galaxy clustering to include geometrical wide-angle effects and other effects that are suppressed under common assumptions, namely, the flat-sky (distant observer) approximation, 
and the approximation that galaxies in a given map are at the same redshift 
(see~\cite{Szalay:1997, Matsubara:1999, Szapudi:2004, Papai:2008, Raccanelli:2010, Raccanelli:2013dza}).

One such effect is sometimes called, in the flat-sky limit, the Doppler term~\cite{Jeong:2011as,Raccanelli:2016doppler}. 
This term is a consequence of the $v/r$ term in the Jacobian relating the real- to redshift-space matter density field,
and is manifested through a combination of the so-called ``mode-coupling'', volume distortions and ``Doppler lensing'' effects.
This additional contribution to RSDs was in fact already present in the seminal paper by Kaiser~\cite{Kaiser:1987}, 
but was first investigated by~\cite{Szalay:1997, Hamilton:1998, Papai:2008}.

Another well-known large-scale effect comes from the primordial non-Gaussianities (pNG) that are imprinted 
at the end of the inflationary era, which leads to a correction to the linear galaxy power spectrum on large scales 
via the halo bias. Namely, during the formation of dark matter halos, small scales are modulated by the 
long-wavelength modes, leading to a scale-dependent bias on very large scales.
This corresponds to adding a scale-dependent correction to the bias which is proportional to the local 
non-linear parameter $f_{\rm NL}$ -- see, e.g., ~\cite{Matarrese:2000iz, Dalal:2007cu, Matarrese:2008nc, Slosar:2008hx, McDonald:2008sc, Desjacques:2010jw, Matsubara:2012nc, Tellarini:2016sgp}, as well as the recent review \cite{Desjacques:2016bnm} and references therein. 

It is well-known that the bias correction from pNG leads to a distortion of the galaxy power spectrum that scales roughly as $\sim k^{-2}$ \cite{Dalal:2007cu}. 
Recently It has been shown that, in the flat sky limit, the Doppler term also induces a 
$\sim k^{-2}$ dependence in the Fourier-space power spectrum~\cite{Jeong:2011as, Raccanelli:2016doppler}, 
which can be degenerate with, and hence potentially a contaminant for, pNG searches. 
As pointed by \cite{Jeong:2011as}, who studied the Doppler and relativistic terms contributing to the redshift-space
power spectrum, these contributions lead to an {\it effective non-Gaussian bias}. However, the scale and
redshift dependence of this effective bias has a particular signature which is distinct from pNGs, which, as we will see, 
enable us to distinguish the $f_{\rm NL}$ from the Doppler dipole.

The main contribution of this paper is to show how the data of low- and intermediate-redshift surveys of 
different types of galaxies (or, more generally, different tracers) can be combined in a way that improves 
the accuracy of measurements of the dipole feature and help disentangle its degeneracy with the non-Gaussianity parameter
$f_{\rm NL}$.

\vspace{0.3cm}
This paper is organized as follows: 
In Section \ref{sec:alpha} we present the usual prescription for transforming from real space to redshift space, 
beginning with the standard scenario \cite{Kaiser:1987, Hamilton:1998} and including the series of local terms such 
as the Doppler term\footnote{Here we are not considering local terms proportional to the gravitational potential, nor, 
for simplicity, non-local (i.e. integrated along the line of sight) corrections.}, assuming the flat-sky limit and in Fourier space.
In section \ref{sec:mocks}, with the help of mock catalogs of galaxies, we illustrate how the contributions of the 
Doppler term can be detected. In Section \ref{sec:data}, using the multi-tracer technique, we estimate the 
redshift-space power spectra for two tracers and isolate the Doppler contribution.
Section \ref{sec:PNG} is devoted to breaking the degeneracy between Doppler effects and pNG. 
In Section \ref{sec:FM} we compute the Fisher matrix using all the information in our mock survey to extract 
constraints on the Doppler terms, $f_{\rm NL}$, as well as other parameters. We conclude in Section \ref{sec:conclusions}.
Our fiducial model is flat $\Lambda$CDM with $\Omega_m=0.26$ and $h=0.7$.

\section{Large-scale features in the redshift-space galaxy-galaxy 2-point functions}
\label{sec:alpha}
In linear theory, the transformation between real space and redshift space is realized by means of a Jacobian
matrix generated by the mapping between radial distances in the two frames. For each individual galaxy,
this mapping can be interpreted as an apparent shift -- depending on local and intervening matter distributions -- of that galaxy projected 
along the line of sight.
In the standard Newtonian picture, the position of a galaxy becomes then $\vec{r}_s = \vec{r} + v_r \hat{r}$, where $r_s$ and $r$ indicate the radial coordinate along the line of sight in redshift- and real-space, respectively, and the shift is proportional to the radial component of the peculiar velocity 
of the galaxy\footnote{More accurately, it is related with the radial displacement $v_r=\hat{r} \hat{v}/\mathcal{H}$, where $\hat{v}$ is  the peculiar velocity of the galaxy.}. In a fully general relativistic picture, this shift also depends on the contributions from time delay, lensing convergence and magnification.
Conservation of the number of galaxies per unit volume means that the mapping must preserve the counts in a volume element, 
$ r^2 \, dr \, d^2\Omega \, n_g(\vec{r}) = r_s^2 \, dr_s \, d^2\Omega \, n_g^{(s)}(\vec{r}_s)$.
As a consequence, in the linear approximation, the fluctuation of the galaxy counts in redshift space inherits a correction that was first studied in~\cite{Kaiser:1987},
\be
\label{deltas}
\delta_g^{(r_s)} = \frac{n_g^{(s)} - \bar{n}_g^{(s)}}{\bar{n}_g^{(s)}}
=\delta_g + \frac{\partial v_r}{\partial r} + \frac{\alpha \, v_r}{r} \; ,
\ee
where $\bar{n}_g$ is the mean number density of galaxies.
This expression has been widely investigated and used in cosmological analyses -- for a review, see e.g.~\cite{Hamilton:1998}.
Since the velocity term is a gradient, the last term in Eq. (\ref{deltas}) leads to a Doppler, which is usually neglected.
This term is proportional to $\alpha$, which accounts for local variations of matter density and volume-lensing distortions.
In the standard Newtonian framework we have~\cite{Kaiser:1987, Hamilton:1998}:
\be
\alpha_{\rm Newt} = 2 + \frac{d \ln \bar{n}_g}{d \ln r} \, .
\ee
The full Jacobian including the velocity ($\alpha$) term was first investigated in~\cite{Zaroubi:1993, Szalay:1997, Hamilton:1998} and later included in simulation and data analyses in~\cite{Papai:2008, Raccanelli:2010, Samushia:2012}.
Given that future galaxy surveys will probe scales comparable to the cosmological horizon, it has been more recently realized that galaxy clustering modeling needs to be based on gauge-invariant quantities, and therefore a general relativistic formalism has been developed (see e.g.~\cite{Yoo:2009, Challinor:2011, Bonvin:2011, Jeong:2011as, Bertacca:2012}). 
In this context, the real-space position of a galaxy is modified to an {\it observed} one by not only peculiar velocity and Doppler terms, but also lensing, magnification, time-delay and ISW-like effects; for a general investigation of the impact of these terms, see~\cite{Raccanelli:2015GR}, and in particular \cite{Reimberg:2015jma} for the case of wide angular separations.

In a fully relativistic framework the $\alpha$ term can be written as~\cite{Bertacca:2012}:
\be
\alpha_{\rm GR} = - r(z) \frac{H(z)}{(1+z)} \left[b_e(z) -
1-2\mathcal{Q}(z) +\frac{3}{2}\Omega_m (z)-
\frac{2}{r(z)}\big[1-\mathcal{Q}(z)\big]\frac{(1+z)}{H(z)}\right] \; ,
\ee
where $r(z)$ is the comoving distance\footnote{More accurately, in \cite{Bertacca:2012} 
$\alpha_{\rm GR}$ depends on  $r_s$, but considering that  $|r_s-r|/r \ll 1$, and assuming only the linear 
redshift space distortions, we can take $r_s \simeq r$.} and
\begin{equation}
b_e ({z})=-(1+ {z}) {d\, \ln \, \bar{n}_g^{(s)} \over d \, z} 
\end{equation}
is the evolutionary bias parameter.
The magnification bias parameter ${\cal Q}$ for a magnitude-limited survey is~\cite{Jeong:2011as}:
\begin{equation}
\label{eq:q}
\mathcal{Q} = -{d \ln \bar{n}_g^{(s)} (L) \over d \ln L }\bigg|_{L=L_{\rm lim}} \; , 
\end{equation}
where $\bar{n}_g^{(s)} (L)$ is the mean redshift-space comoving number density of galaxies with luminosities 
$L>L_{\rm lim}$~\cite{Liu:2014}.
It is worth noting that in literature the magnification bias is sometimes denoted $s(z)$, 
and our magnification bias parameter is $\mathcal{Q}=5s/2$.

The general expression for $\alpha$ was derived\footnote{See also \cite{Jeong:2011as} where the authors defined $\mathcal{B}=\alpha f$.} by \cite{Bertacca:2012} and, recently, has been analyzed in~\cite{Raccanelli:2016doppler}, where, through
modeling in configuration, Fourier and $\ell$-space, it was shown that, for low-redshift maps over large fractions of the sky, the Doppler term can become important.

However, these terms are artificially suppressed in the Fourier space power spectrum,  
due to the combination of the flat-sky (plane-parallel) approximation and the reality condition of the Fourier transform,
which remove any mode-coupling effects.
A redshift-space dipole $D$ introduces a distortion in the density contrast of the form 
$\delta^{(s)} (\vec{r}) = (1 + D \, \hat{r} \cdot \vec\nabla) \delta^{(r)} (\vec{r})$, which in Fourier space 
(using the plane-parallel approximation)
becomes $\tilde\delta^{(s)} (\vec{k}) = (1 + D \, i \, \mu) \tilde\delta^{(r)} (\vec{k})$. Since 
$|\tilde\delta^{(s)}|^2 = (1 + D^2 \mu^2) |\tilde\delta^{(r)}|^2$, the dipole is only manifested in the power spectrum
through a quadrupole $\propto D^2$. Therefore, the power spectrum analysis presented in this paper 
represents a conservative lower bound for detection compared with techniques which keep the 
direct contribution of the dipole. 
Additional reasons for working in $k$-space are the simplicity of covariance 
matrix calculation, and the immediate comparison with local non-Gaussianity effects on large scales, which 
we will show to be partially degenerate with velocity effects.

In a full treatment of the 2-point galaxy correlation function (i.e., including all wide-angle effects), 
the odd Legendre multipoles can serve to distinguish between corrections induced by relativistic 
effects and pNG \cite{Raccanelli:2013dza,Bonvin:2013ogt, Bonvin:2015kuc, Gaztanaga:2015jrs}.
Interestingly, this complete treatment of redshift-space distortions can be extended to 
Fourier space, without the assumption of the plane-parallel approximation -- see \cite{Reimberg:2015jma}.

We start with the full expression at linear order \cite{Challinor:2011, Jeong:2011as} (which includes relativistic effects and magnification bias), and neglect both non-local terms (which, in principle, could contaminate the signal through, e.g., 
the convergence term) as well as local potential terms (which are subdominant compared with velocity terms for distances small compared with the horizon scale $1/H_0$). The result is
the Fourier transform of the density contrast (in the flat-sky limit) in redshift-space, which turns out to be 
proportional to the so-called Kaiser term. Hence, in these approximations 
the redshift-space power spectrum of a galaxy type $i$ can be written as:
\begin{equation}
\label{eq:pk_a}
P^s_i(k,\mu) = \left[ \left(b_i + f \, \mu^2 \right)^2 + \left( \frac{f  \alpha_i}{kr} \mu \right)^2 \right] \, P_m(k) \, 
\end{equation}
where $b_i$ is the bias, $f(z)$ is the matter growth rate, $\mu = \hat{k} \cdot \hat{r}$ 
is the cosine with the line of sight, and $P_m(k)$ is the configuration-space matter power spectrum.
The velocity term $f \alpha_i/kr$ is the Fourier space manifestation of the velocity term $\alpha_i v_r/r$ of Eq. (\ref{deltas}),
with $r(z)$ the (comoving) distance between the observer and the observed volume~\cite{Raccanelli:2016doppler}.

We should stress once again the point that many potentially interesting effects have been neglected in this expression: 
e.g., non-linearities  (through scale-dependent bias, mode coupling, intra-halo velocity dispersion), 
relativistic potential terms, as well as line-of-sight integrated effects~\cite{Raccanelli:radial}.
In addition, by virtue of the flat sky approximation, all odd-parity multipoles vanish by definition (from the
power spectrum, not necessarily from the Fourier transform of the density contrasts), while they are in principle non-zero in a full 3D analysis~\cite{Raccanelli:2013dza}.

The power spectrum of a galaxy type $i$ can be expanded as a sum of multipoles:
\begin{equation}
\label{eq:Pk_alpha}
P_i^s(k,\mu) = \sum_{\ell=0,2,4} P_m^r(k) \, A_i^{(\ell)} (k) \, \mathcal{L}_\ell(\mu) \, ,
\end{equation}
where $\mathcal{L}_\ell$ are Legendre polynomials.
For different tracers of large-scale structure, the bias of each tracer is different, but the velocity field 
(manifested through the matter growth rate) is the same. In addition, the Doppler term can also be
different for each tracer. Hence, the monopole and quadrupole amplitudes are given by:
\begin{eqnarray}
A^{(0)}_i &=&  b_i^2 + \frac{2}{3} b_i f + \frac{1}{5}f^2 + \frac{ f^2 \alpha_i^2}{3k^2 r^2}  \, ; \\
A^{(2)}_i &=& \frac{4}{3}b_i f + \frac{4}{7}f^2 + \frac{2 f^2 \alpha_i^2 }{3k^2 r^2}  \, , \nonumber
\end{eqnarray}
where $b_i$ and $\alpha_i$ are, respectively, the bias and the Doppler amplitude of the tracer species $i$.

In general, the bias of each tracer species can be any function of scale, or redshift.
Assuming the bias and the growth factor to be scale-independent, and in the absence of pNG,
the $\alpha$ term is the only part that induces a scale-dependence. 
It is immediately apparent that this scale-dependence has the same behavior of the one induced by 
primordial non-Gaussianities (pNG) through the squeezed limit of the bispectrum. We will investigate this issue in more 
detail in Section~\ref{sec:PNG}.

Here we investigate the detectability of velocity terms by using simulations of nearly complete, 
low- and intermediate-redshift, very large-area galaxy surveys such as the future J-PAS~\cite{2014arXiv1403.5237B} 
and SPHEREx~\cite{2016arXiv160607039D} surveys.
The best possible accuracy for a measurement of the power spectrum in some Fourier bin is given by
\cite{FKP,Abramo:2012}:
\be
\label{Eq:EffVol}
\frac {\sigma^2[P_i(\vec{k})]} {P_i^2(\vec{k})} \geq \frac{2}{V \, \tilde{V}_{\vec{k}}} \, 
\left[ \frac  {1+ \bar{n}_i P_i(\vec{k})} {\bar{n}_i P_i(\vec{k}) } \right]^{2}  \; ,
\ee
where $\vec{k}$ is a bin in Fourier space, $\bar{n}_i$ is the mean number density of the tracer,
$V$ is the survey volume, and $\tilde{V}_k$ is the volume of the bin in Fourier space.
E.g., a spherical Fourier bin of radius $k$ and width $\Delta k$ has a volume
$\tilde{V}_k = (2\pi)^{-3} \, 4 \pi \,  k^2 \, \Delta k$, while a Fourier bin with azimuthal symmetry $\{ k,\mu \}$ has volume 
$\tilde{V}_{k,\mu}= (2\pi)^{-3} 2 \pi \, \Delta \mu \, k^2 \, \Delta k$, and so on and so forth.
Since the term inside square brackets in Eq. (\ref{Eq:EffVol}) is always greater than one,
the relative uncertainty in the measurement of the spectrum in a bandpower, for a fixed (finite) 
survey volume, is bounded from below -- a consequence of cosmic variance.

Let us point out that Eq. (\ref{Eq:EffVol}) follows from the assumption that the covariance of the power spectrum 
is Gaussian and diagonal in the wavebands. 
However, the survey window and other systematics, which couple different wavelengths and are therefore 
particularly important for the large scales we are interested in, 
are not taken into account in this approximation. In order to take proper care of these real-life issues 
we have simulated full mock maps which reproduce approximately the geometry of each redshift slice of the survey. 
The power spectra are derived from the simulated data just as they would be for real data,
hence the sample variance we obtain for our mock spectra should be 
suitable to represent the data covariance of a large-area galaxy survey.

\section{Mocks}\label{sec:mocks}

In order to illustrate how the contributions of the Doppler term can be detected (and disentangled from 
pNGs) we have simulated a large area ($20.000$ deg$^2$) galaxy survey 
up to $z=0.8$. We generate mock catalogs for two different types of tracers: 
``blue'' galaxies (type=1), with bias $b_1=1.0$
and ``red'' galaxies (type=2), with bias $b_2=2.0$.
Since we use a linear biasing model, it would be more accurate to characterize these tracers as
halos of two different mass ranges -- say, field galaxies (blue galaxies), and groups and clusters (red galaxies). 

We split the survey into four redshift slices, centered on $z=0.1$, 0.3, 0.5 and 0.7, with widths $\Delta z=0.2$.
Each slice is replicated by a box with a different geometry, corresponding approximately to
the transverse areas and radial distances of each slice
-- e.g., the second redshift slice is a $156 \times 156 \times 92$ box with cubic cells of volume 
$\Delta V = (10 \, h^{-1} \, {\rm Mpc})^3$,
yielding a total volume for that slice of $V_2 = 2.24 \times 10^9$ Mpc$^3$.
Each redshift slice is also assigned a mean comoving distance $r_{central, \, i}$, 
in order to implement the Doppler dipole for that slice -- see Eq. (\ref{eq:pk_a}).
The main properties of the tracers, along with the physical characteristics of the redshift slices, are presented in 
Table I.

\begin{table}
\begin{center}
\begin{tabular}{|c||cccc||cc|}
\hline
Slice & $z_{min}-z_{max}$ & $V (h^{-3} \, {\rm Gpc}^3)$ & $r_{central} (h^{-1} \, {\rm Mpc})$ & $f = \Omega_m^{0.55}(z)$
& $\bar{n}_1 (h^3 \, {\rm Mpc}^{-3})$ & $\bar{n}_2 (h^3 \, {\rm Mpc}^{-3})$ \\
\hline
$z_1$ & 0.0-0.2 &	 0.39 & 293 & 0.543 & 	0.11 &  0.02 \\
$z_2$ & 0.2-0.4 &	 2.24 & 839 & 0.643 &	0.05 &  0.01 \\
$z_3$ & 0.4-0.6 &	 4.9 & 1328 & 0.724 & 	0.015 & 0.003 \\
$z_4$ & 0.6-0.8 &	 7.3 & 1765 & 0.786 &	0.012 & 0.002 \\
\hline
\end{tabular}
\caption{Left: redshift slices, volumes in each slice (assuming an area of $20.000$ deg$^2$), mean comoving 
distance to the slices, and matter growth rate $f$ (assumed constant inside each slice). 
Right: mean number densities of the three tracers. The biases ($b_1=1.0$, $b_2=2.0$) 
and amplitudes of the Doppler ($\alpha_1=4.0$, $\alpha_2=2.0$) of 
each tracer are kept fixed on all slices.}
\end{center}
\label{tab:vols}
\end{table}

Each redshift slice was simulated $300$ times through mock galaxy maps based on lognormal density fields
\cite{ColesJones91}. 
Lognormal mocks with galaxies assigned by means of a Poisson random process are a fast, empirical way 
to replicate the statistical properties of the galaxy distribution on large 
scales~\cite{2001ApJ...561...22K,2016MNRAS.459.3693X}.
Each mock corresponds to a different realization 
of Gaussian initial conditions, with a variance that was calibrated such that the
underlying matter power spectrum of the mocks reproduce (on average) the spectrum corresponding
to the Planck satellite's best-fit parameters \cite{planck2015par}, as well as the RSD model of Eq. (\ref{eq:pk_a}).
The density contrast of each galaxy mock is defined in terms of (biased) Gaussian fields $\delta_{G,i}$,
by $\delta_{{\rm LN},i} (\vec{x}) = \exp{ [ \delta_{G,i} (\vec{x}) - \sigma_{G,i}^2/2]} -1$, where 
$\sigma_{G,i}^2 = \langle \delta_{G,i}^2 \rangle$ are the
variances computed over the cells of the grid where the mocks are defined. This assignement 
ensures that the lognormal density fields are bounded from below 
by $\rho_{{\rm LN},i} \geq 0$ (since $\delta_{{\rm LN},i} \geq -1$).
The procedure is also designed in such a way that the lognormal density contrast 
is endowed with correlations inherited through the constraint that
$\tilde\delta_{LN,i} (\vec{k})  = (b_i + f \mu + i f \alpha_i \mu/kr) \tilde\delta_m (\vec{k}) $, with 
$\langle \tilde \delta_m (\vec{k}) \tilde\delta_m (\vec{k}{\,}') \rangle = (2\pi)^3 \delta_D(\vec{k}-\vec{k}{\,}') P_m(k) $.
Galaxies are introduced at the last stage, by assigning each cell on the grid with 
a random number of galaxies taken from a Poisson distribution with expectation value $\bar{N}[1+\delta_{LN}(\vec{x})]$,
where $\bar{N}$ is the mean number of galaxies per cell.

A clarification is in order at this point. As was outlined in Section \ref{sec:alpha},
we plan to study redshift-space distortions in simulated galaxy maps. 
However, on large scales we should be careful to apply the general-relativistic approach 
\cite{Yoo:2009, Challinor:2011, Bonvin:2011, Jeong:2011as, Bertacca:2012}. 
With that in mind, the correct definition of bias, i.e. the gauge-invariant bias, is equal to the linear bias relation between 
$\delta_g$ and $\delta_m$ in the synchronous comoving gauge \cite {Jeong:2011as}.
But the matter density contrast in the synchronous comoving gauge is 
equivalent to the comoving contrast in terms of number counts (if we neglect the late-time effects of radiation 
perturbations -- see, e.g. see \cite{Chisari:2011iq,Borzyszkowski:2017ayl}), 
which is in the prescription used both in Newtonian N-Body simulations and in our lognormal mocks.

We extract the power spectra from each mock, and deconvolve the window function from the 
estimation in order to obtain the ``observed'' power spectra -- see Section \ref{sec:data} for details.
The multipoles of the power spectra were computed using the method proposed in \cite{2015PhRvD..92h3532S},
which employs the physical multipoles of the maps (computed in configuration space). 
In contrast, the usual approach assumes a fixed 
line-of-sight to the center of the survey (e.g., $\hat{r} \to \hat{z}$), which is then used to define the orientation of the 
Fourier modes ($\mu \to \hat{k} \cdot \hat{z}$).
For large-angle effects in large-area surveys, fixing a line-of-sight is particularly problematic and could lead
to large systematic effects -- which is the reason we prefer to follow \cite{2015PhRvD..92h3532S}.

In addition to the 1200 mocks including the dipole contributions ($\alpha_1=4.0$, $\alpha_2=2.0$), we also created
1200 mocks without the dipole contribution, so we could run a simple null test for a first check of the detectability 
of the dipole term by computing $\Delta P_i^{(\ell)} = P^{(\ell)}_{i,{\rm dip}} - P^{(\ell)}_{i,{\rm No \, dip}}$. 
Since the sample covariance of the power spectra with and without the dipole term are nearly identical, 
and the dipole term is a small correction to the power spectra (even on the largest scales), it is a good approximation to
employ that sample covariance as the uncertainty for $\Delta P_i^{(\ell)}$.
This means that we can estimate the detectability of the Doppler term by computing the inverse signal-to-noise ratio
$\Delta P_i^{(\ell)} /\sigma(P_i^{(\ell)})$.

\begin{figure}
\resizebox{7.5cm}{!}{\includegraphics{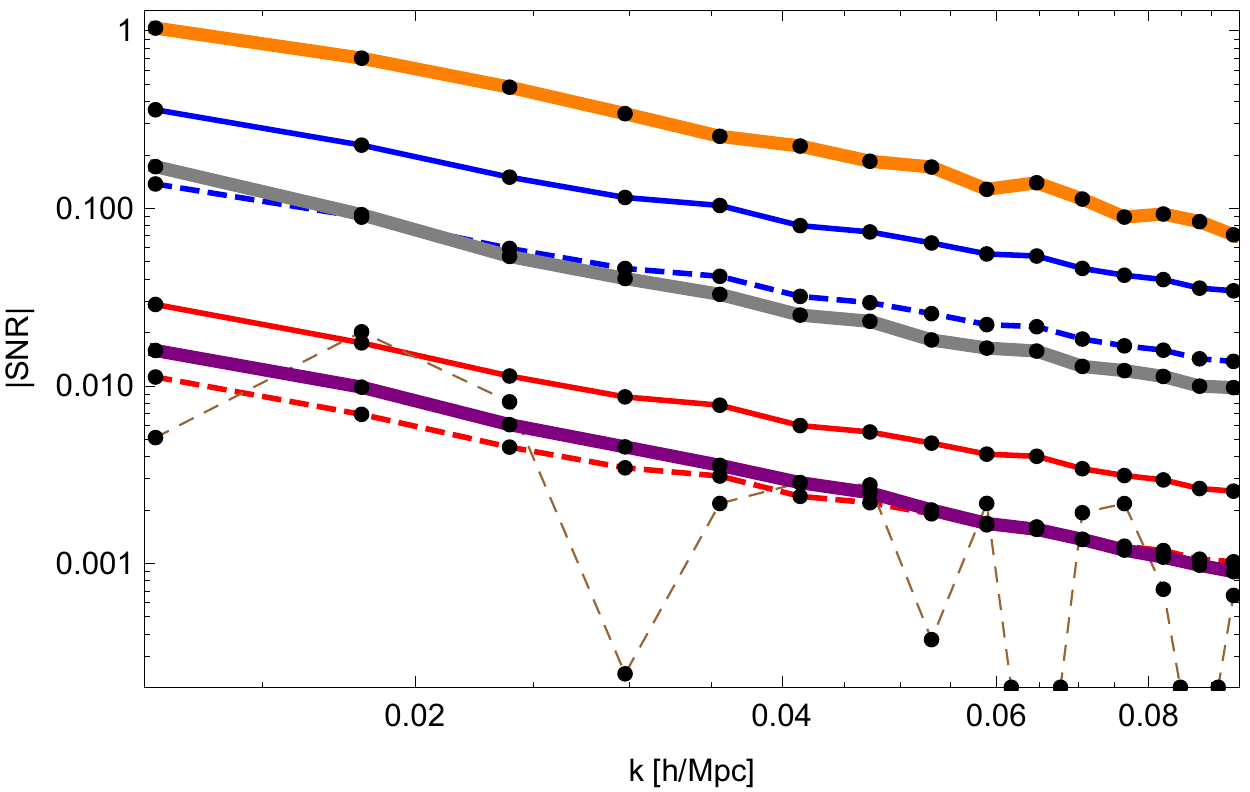}}
\hskip 0.5cm
\resizebox{7.5cm}{!}{\includegraphics{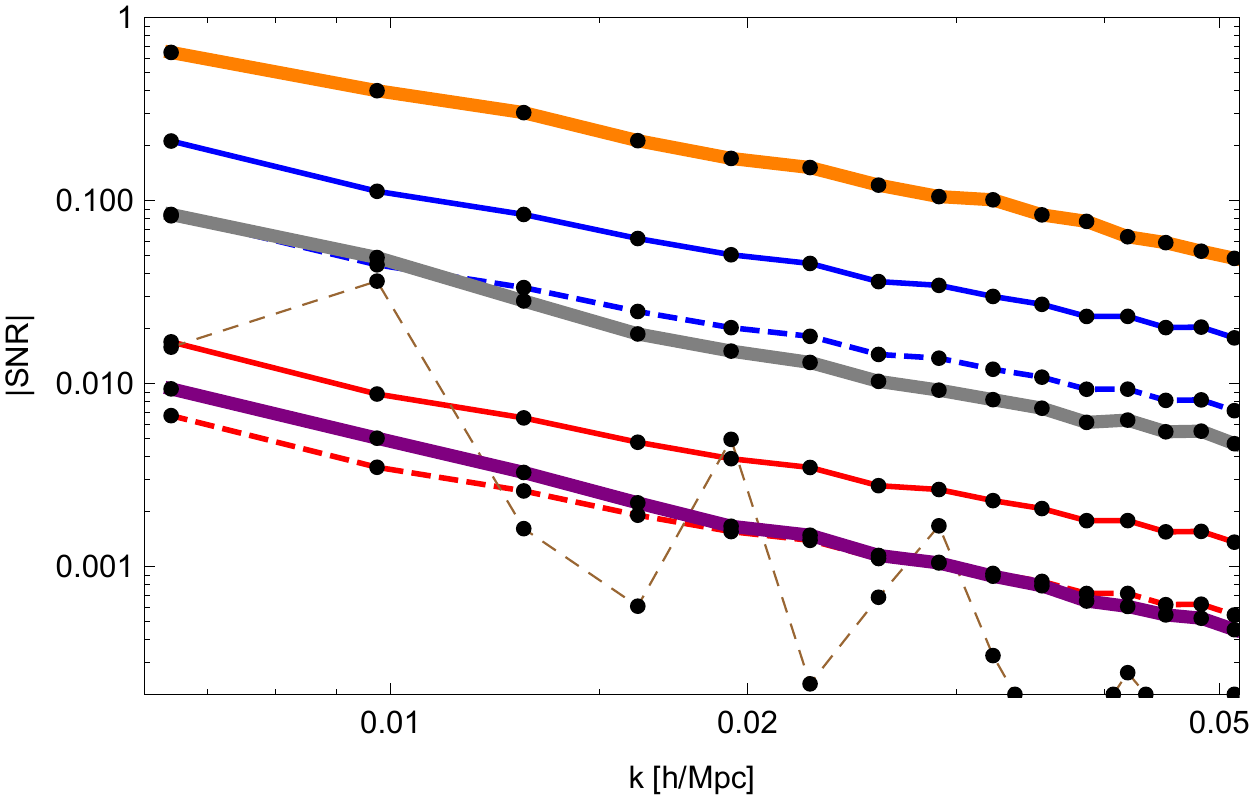}}
\resizebox{7.5cm}{!}{\includegraphics{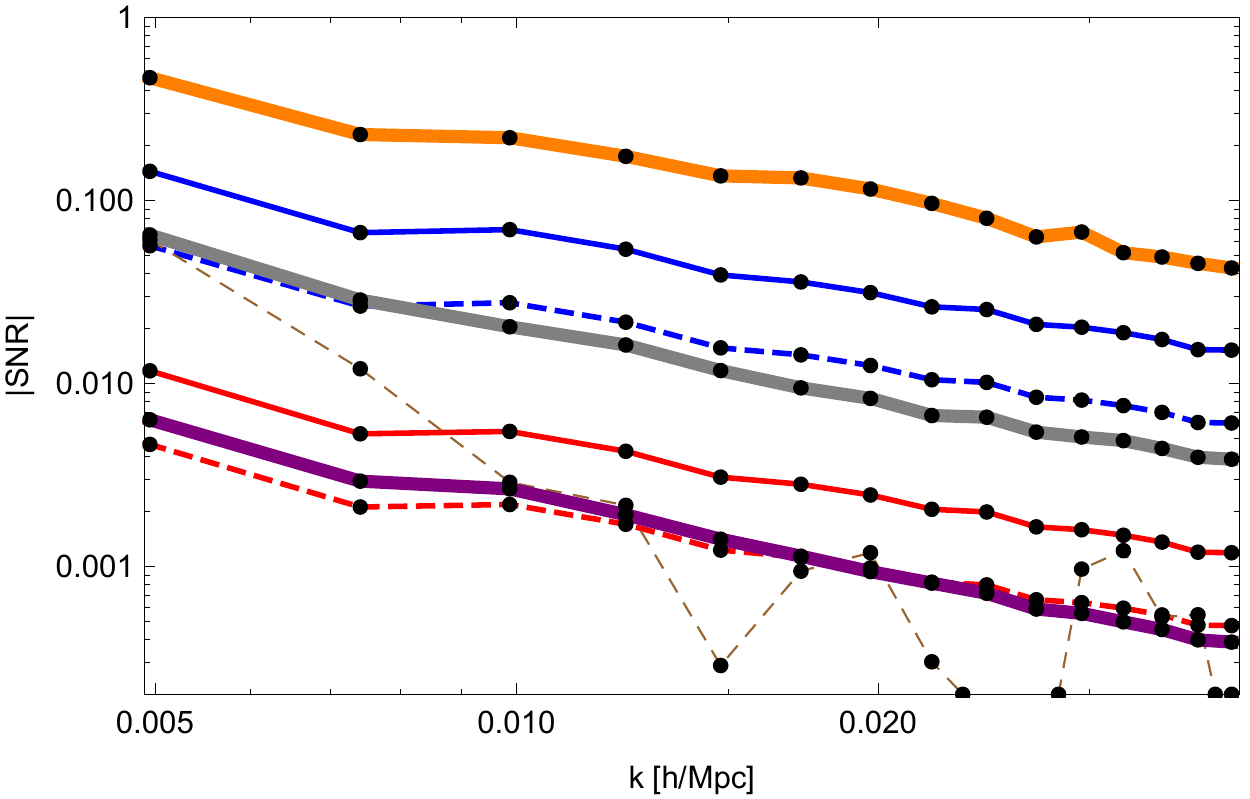}}
\hskip 0.5cm
\resizebox{7.5cm}{!}{\includegraphics{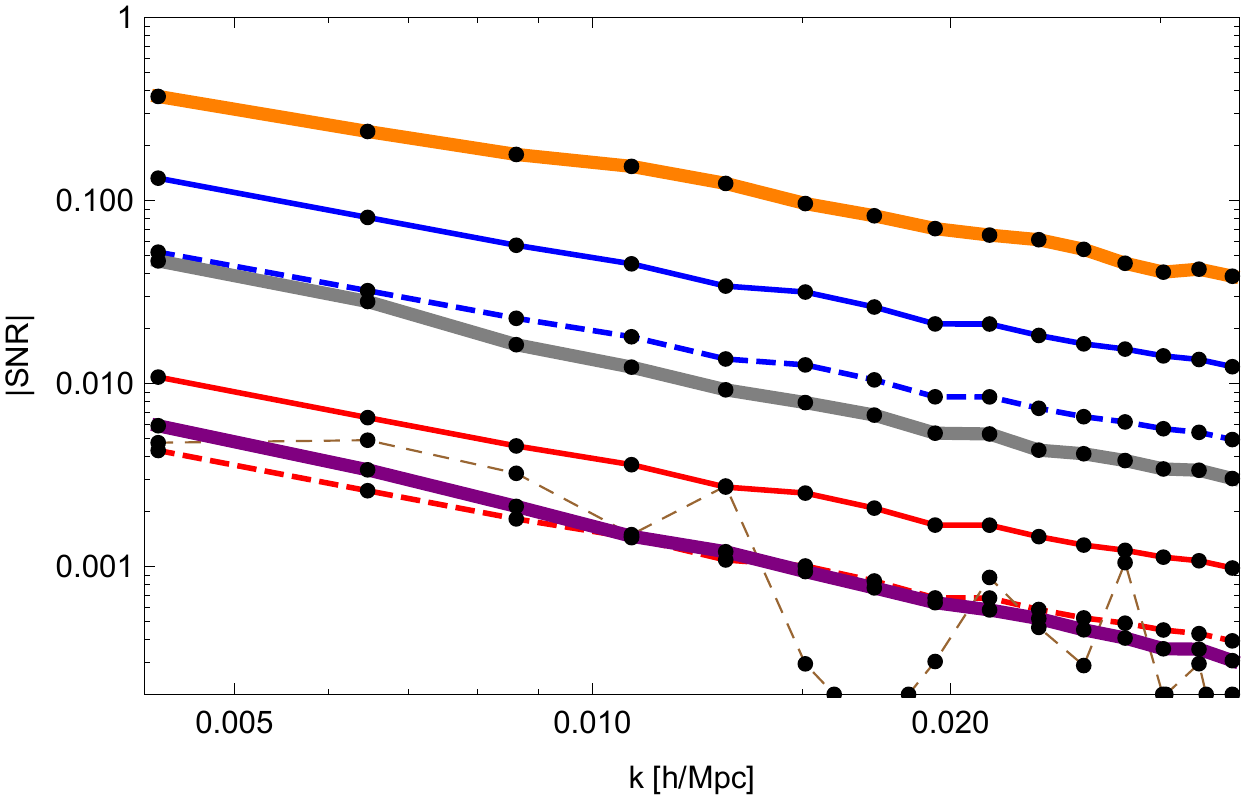}}
\caption{Signal-to-noise ratios (SNR) for different observables, defined as $SNR = | \Delta X|/\sigma(X)$ .
Thin lines (solid for the monopoles; dashed for the quadrupoles) indicate the SNRs for the 
monopole and quadrupole of ``blue galaxies'' (tracer=1, blue in color version) and ``red galaxies'' 
(tracer=2, red in color version). The four panels correspond to the redshift slices around $z=0.1$, 0.3, 0.5 and 0.7, 
with $\Delta z=0.2$ -- see Table 1 for more details.
The thick lines, from top to bottom, correspond to the ratios
$P^{(0)}_1/P^{(0)}_2$ (orange in color version), $P^{(2)}_1/P^{(0)}_1$ (gray in color version), and 
$P^{(2)}_2/P^{(0)}_2$ (purple in color version). The ratios of the quadrupoles of the two tracers (thin dashed lines) 
are also shown for completeness, but the SNR in that case has lower significance.}
\label{Fig:SNRs}
\end{figure}

In Fig.  \ref{Fig:SNRs} we show the signal-to-noise ratios (SNRs) for the detectability of the Doppler term,
for a variety of observables, in a limited range of bandpowers. 
Thin lines (solid for the monopoles; dashed for the quadrupoles) indicate the SNRs for the 
monopole and quadrupole of the galaxy populations $1$ (blue in color version) and $2$ (red in color version), i.e., 
$\left| \sigma[P^{(\ell)}_i(k)]/P^{(\ell)}_i(k) \right|$. The thick lines, from top to bottom, correspond to the ratios
$P^{(0)}_1/P^{(0)}_2$ (orange in color version), $P^{(2)}_1/P^{(0)}_1$ (gray in color version), and 
$P^{(2)}_2/P^{(0)}_2$ (purple in color version). The ratio of the quadrupoles of the two tracers (thin dashed line) 
is also shown for completeness, but its SNR is very small and strongly affected by sample variance. 
Apart from the ratio of quadrupoles, the distributions of the observables shown in this figure are very nearly Gaussian.

We can also compute a ``null test'' for the detectability of the dipole term by means of a chi-squared:
\begin{equation}
\label{Eq:Chi}
\chi^2_{i (\ell)} = \sum_k \left[ \frac{\Delta P_i^{(\ell)} (k)}{\sigma(P_i^{(\ell)})} \right]^2 \; .
\end{equation}
For the lowest-redshift slice ($z=0.1$), using only the monopole of the spectrum of the first tracer 
(``blue galaxies'') results in a $\chi^2_{1 (0)} = 0.26$, while using only the monopole of the second tracer results in
an almost insignificant signal at $\chi^2_{2 (0)} = 0.03$. The quadrupoles of either tracers lead to even lower significance.

Hence, the $p$-value for rejecting the null test using only the amplitudes of the power spectra is at best
around 0.39 (the cumulant for a $\chi^2=0.26$ with a single degree of freedom). However, it is clear from Fig. \ref{Fig:SNRs}
that the ratio of the monopoles of the two tracers has a much higher significance, which should translate in
an improved potential for detecting the effects of the Doppler.
We will show in the following sections how it will be possible to make this detection, and how it
can be disentangled from pNGs.

\section{Estimation of the Doppler dipole term}
\label{sec:data}

\subsection{The multi-tracer technique}

In any given cosmic volume the tracers of the large-scale structure are bound to follow the same 
dark matter distribution in that particular volume. 
Different galaxies may have different biases, but in the linear approximation their 
velocities are driven by the same gravitational potential, hence their power spectra must be 
proportional to $P_m(k)$. 
It was first pointed out by \cite{Seljak:2008, McDonald:2008} that cosmic variance, which is a key
source of variance for the matter power spectrum, should cancel out when taking ratios of the power spectra of
different tracers. 
Hence, surveys that distinguish different types of tracers (i.e., tracers with different biases) allow us to 
measure ratios of the clustering properties of those tracers, to an accuracy which is not limited by 
cosmic variance \cite{Abramo:2012,Abramo:2013}.

Whereas the amplitude of the galaxy power spectrum has an uncertainty which is at best the one given
in Eq. (\ref{Eq:EffVol}), a measurement of the ratio of the power spectra of two
different tracers, $f_{12} (\vec{k}) = P_1 (\vec{k})/P_2 (\vec{k})$, has an uncertainty which is bound by:
\be
\label{eq:errf}
\frac{\sigma^2[f_{12}(\vec{k})]}{f_{12}^2(\vec{k})} \geq \frac{4}{V \, \tilde{V}_{\vec{k}}} \, 
\frac {1+ \bar{n}_1 P_1(\vec{k}) \, + \, \bar{n}_2 \, P_2(\vec{k})}  {\bar{n}_1 P_1(\vec{k}) \, \, \bar{n}_2 \, P_2(\vec{k})} \; ,
\ee
and similar expressions when there are more than two tracers \cite{Abramo:2013,Abramo:2016}.
This means that, in contrast to the uncertainties that apply for the amplitudes of the individual power spectra, 
ratios of spectra can be measured to an arbitrarily high precision, as long as the tracers are 
sufficiently abudant ($\bar{n}_i P_i \gg 1$).
For many physical observables, in particular those related to RSDs, this means an enhancement of the
constraining power coming from measuring the ratios of the power spectra of different types of tracers.
Moreover, since the power spectrum peaks around $k \sim 0.02 \, h \, {\rm Mpc}^{-1}$, this condition is more easily 
satisfied on large scales, which is indeed where we will look for the imprints left by the Doppler term and
by pNG.

Cosmic variance does not apply for ratios of spectra because, in the absence of mode-coupling, 
the random process which sets the initial conditions for the cosmic density field cancels out for 
observables such as $P_i(k,\mu)/P_j(k,\mu)$. In other words: even though the power 
spectrum which is realized inside any given (finite) 
volume is subject to the statistical fluctuations which follow from the (nearly Gaussian) random 
nature of the density field, the ratio of two spectra is not. Indeed, physical processes which 
affect different objects moving in this density field, and which are not degenerate with the statistical 
fluctuations of the density field, should be determined to an accuracy that has nothing to do
with cosmic variance. Hence, galaxy bias, the peculiar velocity field, as well as any other physical effects 
which are manifested by means of different relative clustering strengths for different tracers (such
as the Doppler and pNG),
can be measured to an accuracy that is only limited by the numbers of those tracers which can be effectively
detected \cite{Seljak:2008, McDonald:2008, Abramo:2013}. 

In particular~\cite{Abramo:2013} showed explicitly, by using the multi-tracer Fisher information matrix, 
that ratios of spectra (combined in a certain way) constitute physical observables that are 
uncorrelated with the matter power spectrum itself. 
Given $N$ different tracers of large-scale structure,
there is only one observable which gives the highest signal-to-noise ratio for the power spectrum
(the ``effective power spectrum'' of the survey, also found by~\cite{PVP}), and $N-1$ observables which are
ratios of spectra of the different tracers -- all uncorrelated with each other.

The main limitation of the multi-tracer technique is the biasing of the tracers. 
Although redshift and scale dependences $b(z,k)$ can be easily included and do not change Eq. (\ref{eq:errf}),
mixing tracers of different biases results in a degradation of the signal. 
E.g., in the Halo Model~\cite{CooraySheth}, galaxies with a certain halo occupation distribution (HOD) appear 
randomly in a range of halos of different masses, and therefore their bias reflects the halo biases which 
correspond to that range. 
Unless the ranges of halo masses spanned by the HODs of the two galaxy types considered here 
do not overlap, this stochasticity can smear out the information of the original set of tracers with deterministic 
biases~\cite{Abramo:2016}.

An important step towards realizing the multi-tracer technique was obtained in~\cite{Abramo:2016},
who obtained the optimal weights that should be used to estimate the individual auto-power spectra in a multi-tracer survey.
These multi-tracer weights generalize the traditional weights of~\cite{FKP} (FKP, which are
optimal only for a single tracer), as well as those found by~\cite{PVP} for multiple tracers,
but which only optimize the estimation of the effective power spectrum of the survey,
${\cal P}_{eff} (k) = \sum_i \bar{n}_i P_i (k)$.
In~\cite{Abramo:2016} it was shown that the covariance of the multi-tracer power spectrum
estimators is always either equal to, or smaller than, the covariance matrix obtained through the use of 
the FKP estimator.

\subsection{Estimation of the power spectra}

Given a galaxy survey, the power spectrum can be computed basically in three steps:
(i) take the density contrast of galaxies, $\delta_i=(n_i - \bar{n}_i)/n_i$, and up-weight the denser 
regions of the survey, creating a weighted density field $F_i (\vec{x})$; 
(ii) take the Fourier transform of the weighted density contrast, $\tilde{F}_i (\vec{k})$; 
and (iii) compute the averages $\langle | \tilde{F}_i |^2 \rangle_{\vec{k}} \sim P_i(\vec{k})$,
where $\vec{k}$ here stands for a bin in Fourier space (a bandpower).

In a seminal paper, Feldman, Kaiser \& Peacock \cite{FKP} showed that there are certain weights 
(the FKP weights) which minimize
the covariance of the estimated power spectra, such that, under ideal conditions, that covariance is equal
to the inverse of the Fisher information matrix of the power spectra. Since this is the Cr\'amer-Rao bound,
no estimator has a lower variance compared with the one that employs the FKP weights -- which are optimal, in that sense.

However, the FKP weights are only optimal under the assumption that there is a single type of tracer in the survey, with 
a single bias and a single shape of RSDs. This is never the case: galaxies comes in many different types and biases, at 
different redshifts, and, more fundamentally, they occupy different dark matter halos.

This problem was first tackled by Percival, Verde \& Peacock \cite{PVP}, who showed that, under 
similar assumptions as the ones necessary to derive the FKP weights, the optimal estimator 
for the matter power spectrum in a multi-tracer survey necessitates a different weighting scheme. 
However, the PVP weights were designed to optimize the estimation of a single
parameter: the survey's effective power spectrum ${\cal P}_{eff} (k) = \sum_i \bar{n}_i P_i (k)$ -- not the redshift-space 
power spectra of each individual tracer.

This issue was solved by \cite{Abramo:2016}, who obtained multi-tracer weights 
that are optimal for the estimation of the redshift-space power spectra of an arbitrary number of tracers 
which share the same cosmic volume. Both the FKP and PVP weights are, in fact, projections of the 
multi-tracer weights, which result from combining the individual power spectra of the tracers to
produce aggregate observables such as the effective power spectrum of the survey.
In particular, it was shown in \cite{Abramo:2016} that the multi-tracer weights are always superior
to the FKP weights for the estimation of the individual auto-power spectra of two or more different types of tracers 
of large-scale structure.

The multi-tracer weighted fields are defined as $F_i = \sum_j w_{ij} \delta_j$, where the weights are~\cite{Abramo:2016}:
\be
\label{eq:mtw}
w_{ij} = \bar{n}_i b_i \delta_{ij}  - \frac{\bar{n}_i P_i \, \bar{n}_j b_j}{1+{\cal{P}}_{eff}} \; .
\ee
For a single tracer species this reduces to the FKP weights for the density contrast, $w \to \bar{n} b/(1+ \bar{n} P(k))$.
For two or more distinct tracer species, the weights have the effect of mixing the different tracers, such
that the auto-correlations of the weighted fields $F_i$ include not only the auto-correlations of the tracers, but 
their cross-correlations as well.
The multi-tracer weights combine the different tracers in a manner that optimizes the measurement
of the auto-power spectra $P_i(\vec{k})$, subject to the constraint that any auto- or cross-spectra are 
related to the underlying matter power spectrum by a multiplicative biasing and RSD model, 
namely $\langle \tilde\delta_i (\vec{k}) \tilde\delta_j^* (\vec{k}{\,}') \rangle = 
(2\pi)^3 \delta_D(\vec{k} - \vec{k}{\,}')
B_i(\vec{k}) B_j(\vec{k}) P_m(k)$ \cite{Abramo:2016}.

\begin{figure}
\resizebox{7.5cm}{!}{\includegraphics{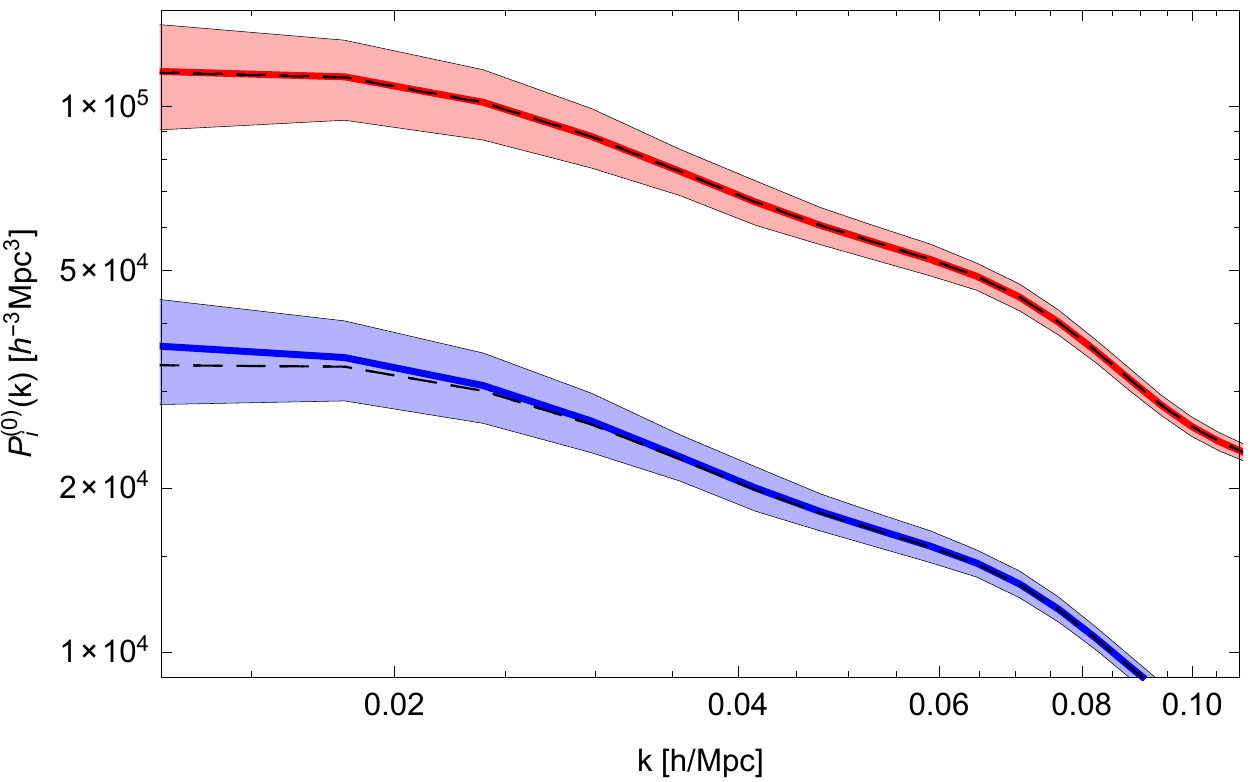}}
\hskip 0.5cm
\resizebox{7.5cm}{!}{\includegraphics{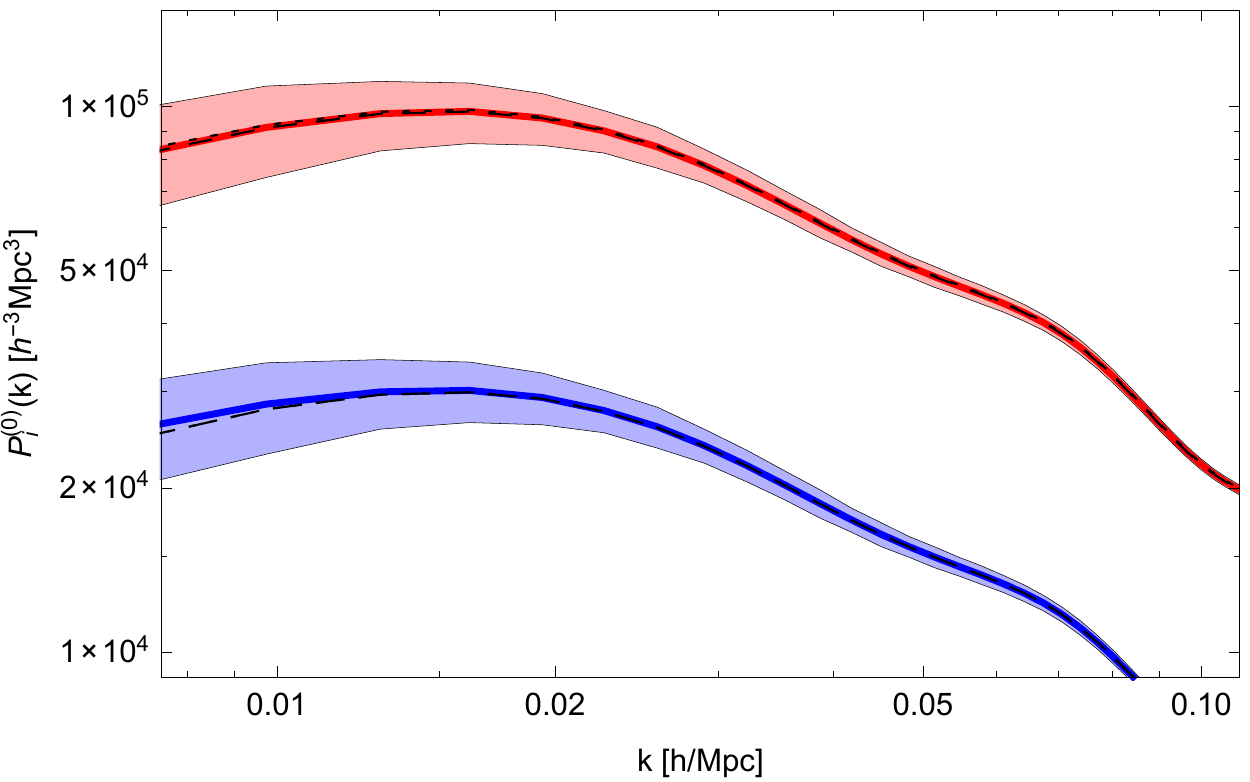}}
\resizebox{7.5cm}{!}{\includegraphics{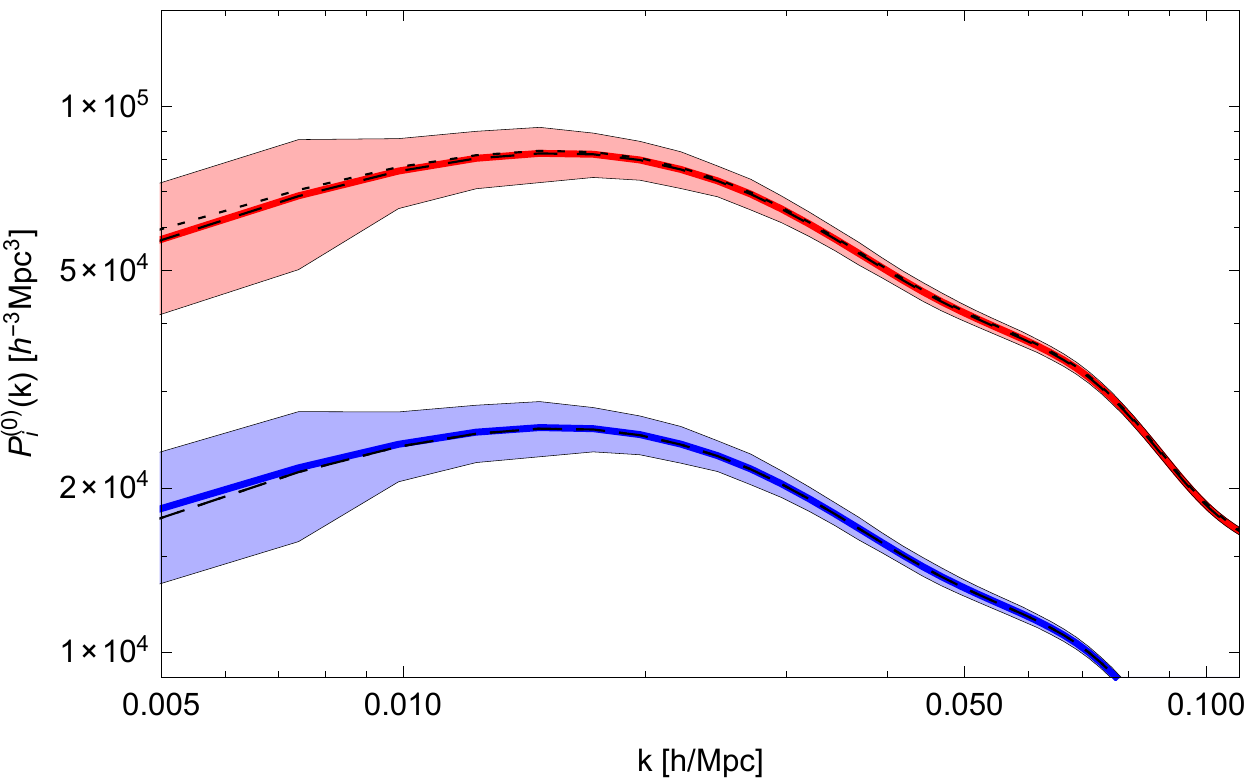}}
\hskip 0.5cm
\resizebox{7.5cm}{!}{\includegraphics{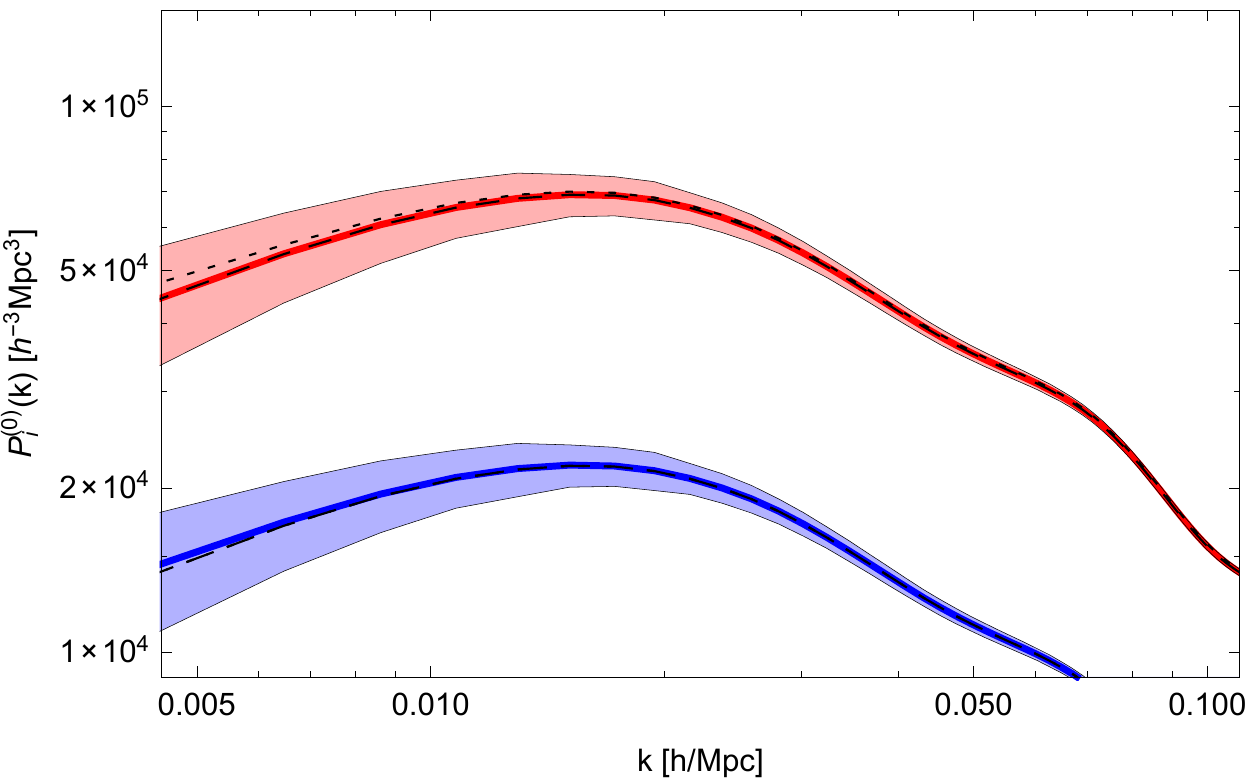}}
\caption{Monopoles of the two tracers used in this work, in each of the four redshift slices (see Table II):
$z_1$ (top left), $z_2$ (top left), $z_3$ (bottom left), and $z_4$ (bottom right).
Solid lines and shaded regions indicate, respectively, the mean and variance of 300 mocks for each slice. 
In each mock the distributions of blue galaxies (tracer type 1) and 
red galaxies (type 2) follow the same realizations of the density field on each mock.
The dashed lines indicate the theoretically expected spectra with a null Doppler term, and
the dotted lines correspond to null Doppler and an 
amplitude of pNGs of $f_{\rm NL}=10$. 
}
\label{Fig:P00_spec}
\end{figure}

\begin{figure}
\resizebox{7.5cm}{!}{\includegraphics{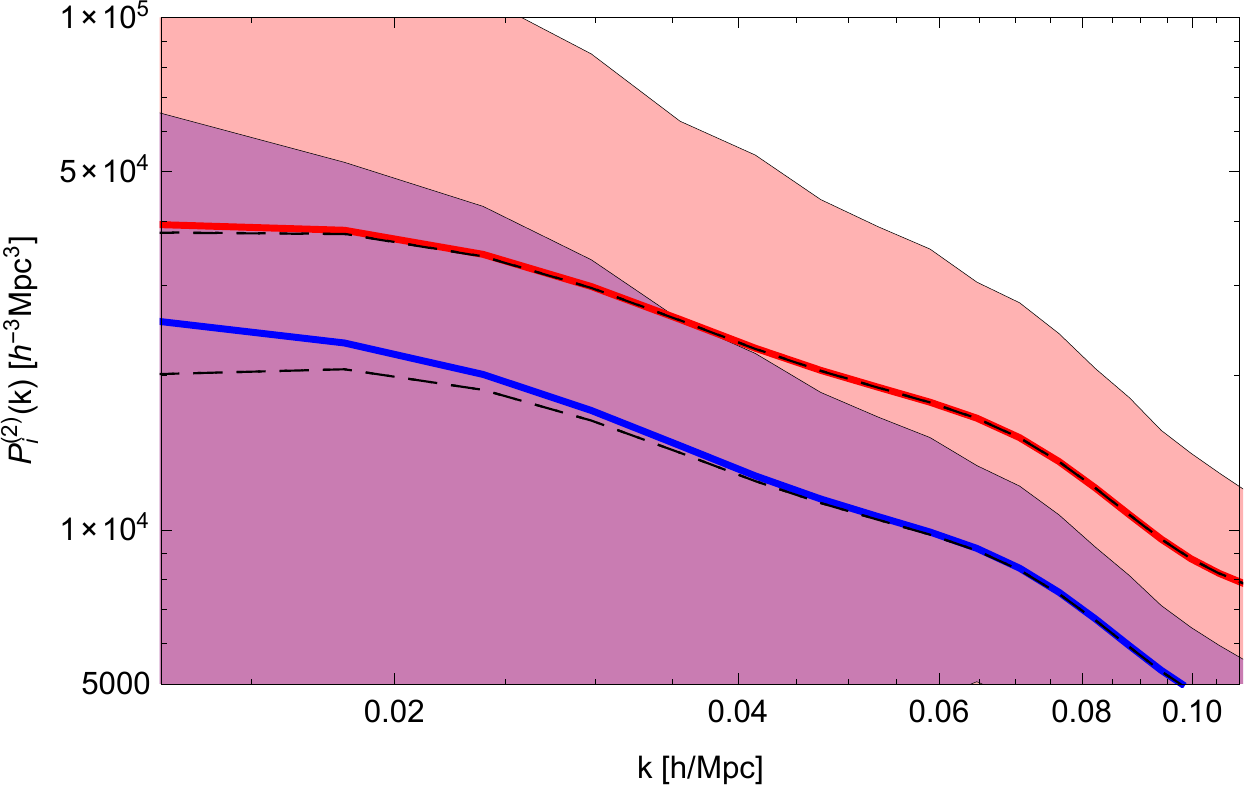}}
\hskip 0.5cm
\resizebox{7.5cm}{!}{\includegraphics{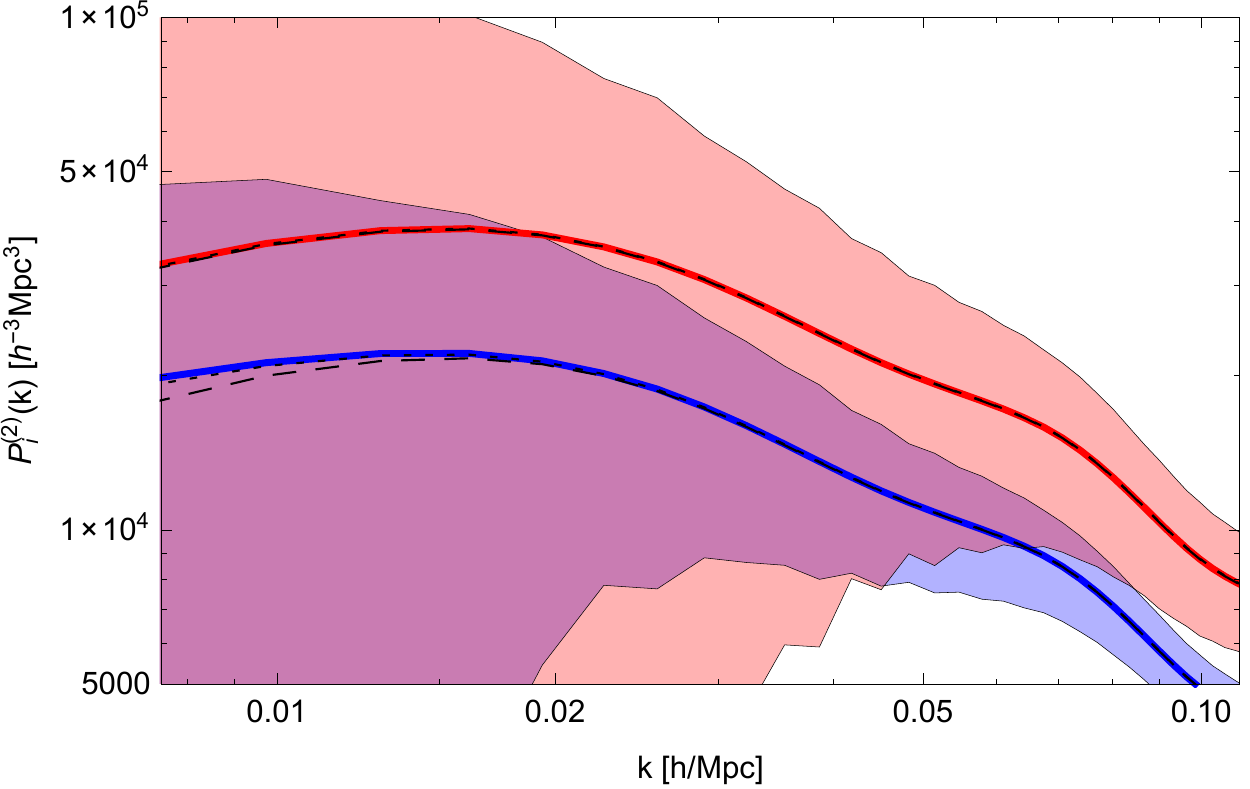}}
\resizebox{7.5cm}{!}{\includegraphics{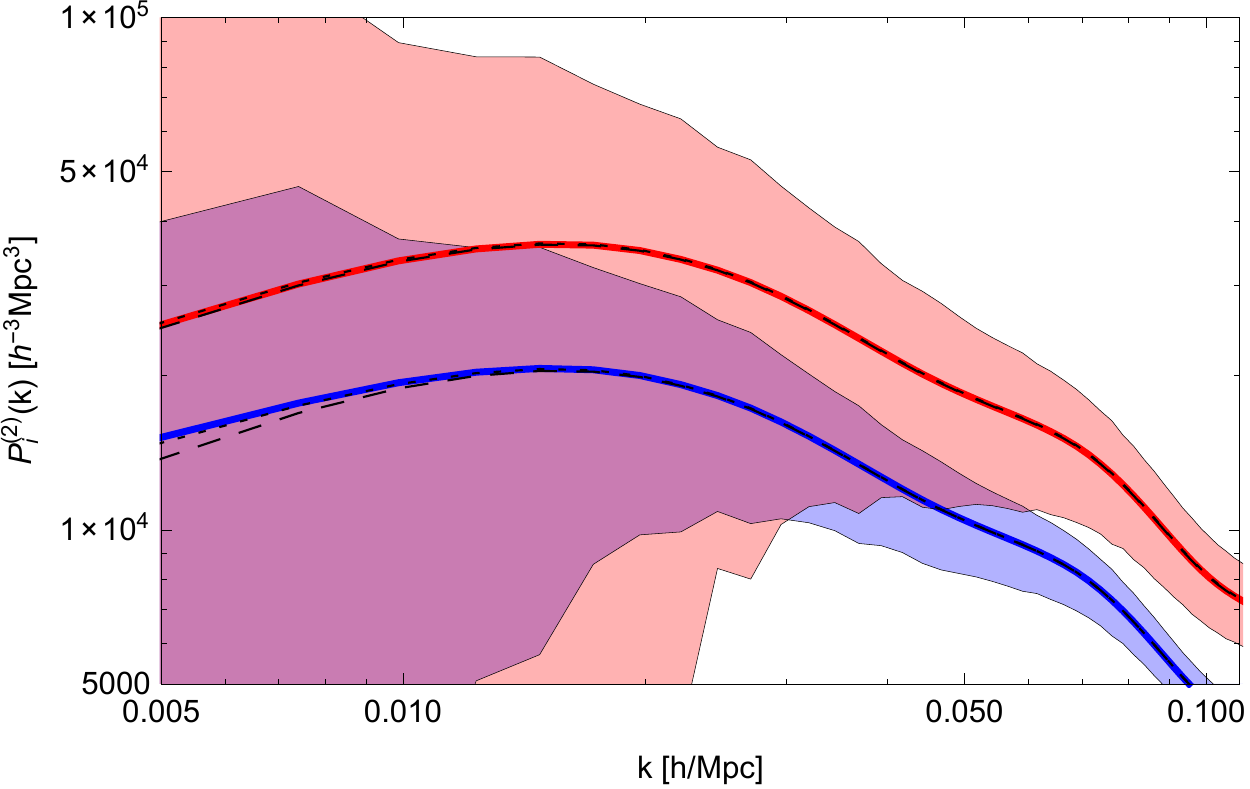}}
\hskip 0.5cm
\resizebox{7.5cm}{!}{\includegraphics{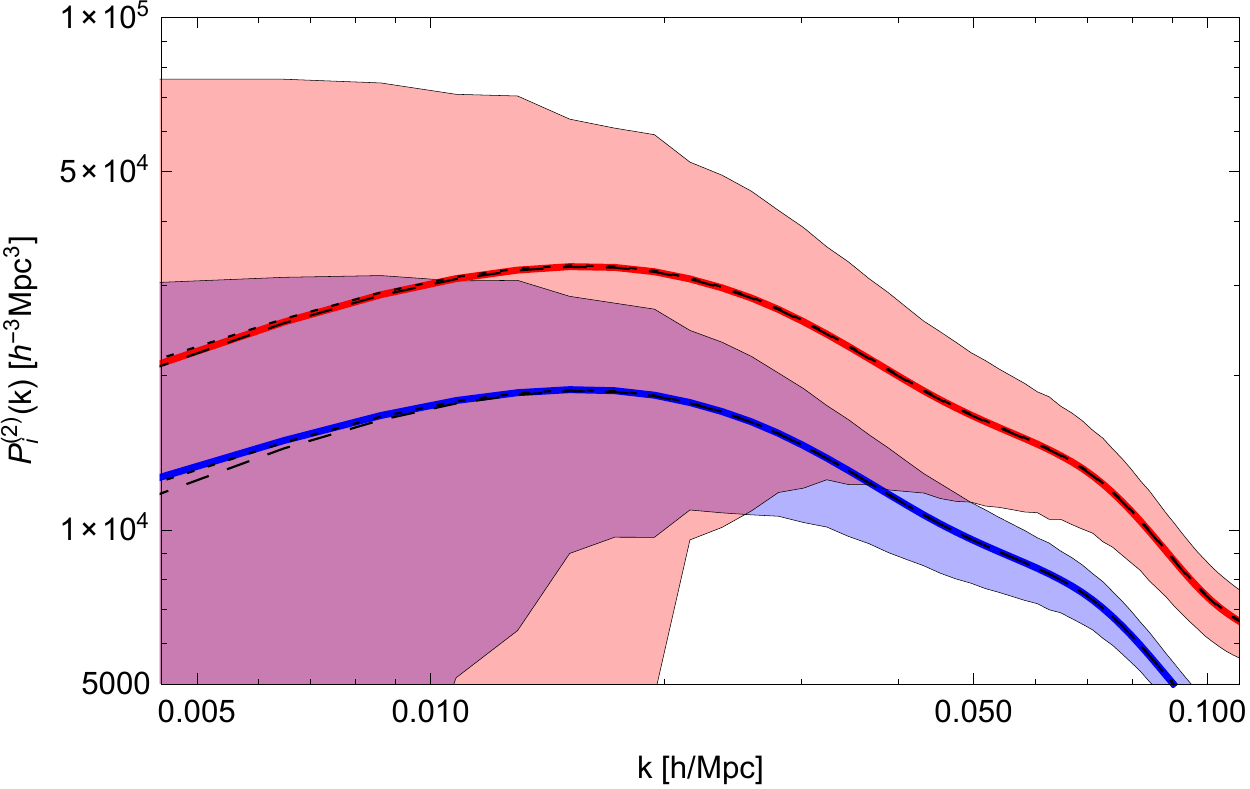}}
\caption{Quadrupoles of the two tracers used in this work. As in Fig. \ref{Fig:P00_spec}, the four redshift slices 
are shown from left to right, top to bottom. The same colors and labels as in Fig. \ref{Fig:P00_spec} apply.
The quadrupole has, naturally, a larger variance compared with the monopole -- na\"ively, $5 \times$ larger.}
\label{Fig:P22_spec}
\end{figure}

In Fig. \ref{Fig:P00_spec} we show the monopoles [$P_i^{(0)}(k)$], and in Fig. \ref{Fig:P22_spec} the quadrupoles 
[$P_i^{(2)}(k)$], of tracer 1 (blue galaxies, lower lines) and of
tracer 2 (red galaxies, upper lines), obtained using the multi-tracer weights. 
From top to bottom, left to right, the panels correspond to the four redshift slices (see Table II). 
Thick lines are the sample mean, and shaded areas indicate sample variances.
We also show both multipoles without the Doppler term (dashed lines), 
as well as the multipoles without the Doppler term but with a level of pNGs of $f_{\rm NL}=10$ (dotted lines).
We have found that the set of multipoles of our sample is very well described by a multivariate Gaussian, so from
now on we only consider their covariance in order to describe the likelihood of the data.

\subsection{Large-angle effects estimators}

Percival and White (2009, henceforth PW) suggested an estimator for the Doppler term which is a linear combination of the
three multipoles ($\ell = 0, 2$ and 4). 
Since the SNR of the $\ell=4$ multipole is negligible, one can think of using just the monopole, $P^{(0)}_i(k)$, 
and the quadrupole, $P^{(2)}_i(k)$. In particular, one can estimate the ratios:
\begin{equation}
\label{mono}
\frac{P_i^{(2)}(k)}{P_i^{(0)}(k)} = 
\frac{\frac{4}{3}b_i f + \frac{4}{7}f^2 + \frac{2\, \alpha_i^2 f^2}{3k^2 r^2}}
{b_i^2 + \frac{2}{3} b_i f + \frac{1}{5}f^2 + \frac{\alpha_i^2 f^2}{3 \, k^2 r^2}} \; .
\end{equation}

However, in a survey with two (or more) tracers of large-scale structure, other estimators can be constructed,
which make use of the different redshift-space power spectra of the tracers.
Indeed, it was first suggested by \cite{Seljak:2008,McDonald:2008} that ratios of power spectra could 
be measured to an accuracy which is not limited by cosmic variance. In \cite{Abramo:2013} it was shown how the
diagonalization of the multi-tracer Fisher matrix leads to these ratios, which represent the statistically independent, 
uncorrelated observables for a multi-tracer cosmological survey.

The derivation of these ratios of power spectra  is rather simple: given the original redshift-space 
power spectra for the $N$ tracers, we first write 
these spectra in units of shot noise ($1/\bar{n}_i$): ${\cal{P}}_i \equiv \bar{n}_i P_i$. 
These variables can be regarded as coordinates in an $N$-dimensional space 
(or, in this case, since the ${\cal{P}}_i >0$, a wedge), 
where the multi-tracer Fisher matrix plays the role of a metric for the information \cite{Abramo:2013} 
\footnote{In fact, that space has $N \times N_{\vec{k}}$ dimensions, where $N_{\vec{k}}$ is the number 
of Fourier bins (bandpowers). However, for large enough bins, there are no correlations between different bins,
and the multi-tracer Fisher matrix becomes block-diagonal \cite{Abramo:2012}.}.
Now, identify the spectra with (squared) Cartesian coordinates in this space, ${\cal{P}}_i \to x_i^2$.
The coordinate transformation that diagonalizes the multi-tracer Fisher matrix consists in changing from Cartesian
to spherical coordinates, with the radial coordinate being $r^2 = \sum_i x_i^2$, and the
$N-1$ angles given by the usual formulas for their tangents \cite{Abramo:2013}. 
It is irrelevant which tracer species plays the role of the $z$ direction (the polar axis): 
any choice results in a valid spherical coordinate system, and results in 
the diagonalization of the Fisher matrix.

As an example, in a survey with three galaxy types we could choose 
${\cal{P}}_1 \to x^2$, ${\cal{P}}_2 \to y^2$ and ${\cal{P}}_3 \to z^2$, 
which leads to the two ratios of spectra corresponding to the 
tangents of the azimuthal and polar angles, as in:
\bea
\label{ts}
t_1 = \frac{\bar{n}_2 P_2}{\bar{n}_1 P_1} 
\quad &\Leftrightarrow& \quad 
\tan^2 \varphi = \frac{y^2}{x^2} \; , 
\\ \nonumber
t_2 = \frac{\bar{n}_3 P_3}{\bar{n}_1 P_1 + \bar{n}_2 P_2} 
\quad &\Leftrightarrow& \quad 
\tan^2 \theta = \frac{z^2}{x^2+y^2} \; .
\eea

Together with ${\cal{P}} _{eff}= {\cal{P}}_1 + {\cal{P}}_2 + {\cal{P}}_3$ (the radial variable), 
the observables $t_1$ and $t_2$ form a complete
and uncorrelated set of physical observables for a survey consisting of three tracers of large-scale structure. 
In 3D Fourier space, the two observables $t_1 (k,\mu)$ and $t_2 (k,\mu)$ carry all the information 
that can be extracted from the ratios of the spectra of the three tracers, while ${\cal{P}}$ (which is
precisely the survey's effective power spectrum of \citet{PVP}) carry all
the information about the matter power spectrum. 

\begin{figure}
\resizebox{7.5cm}{!}{\includegraphics{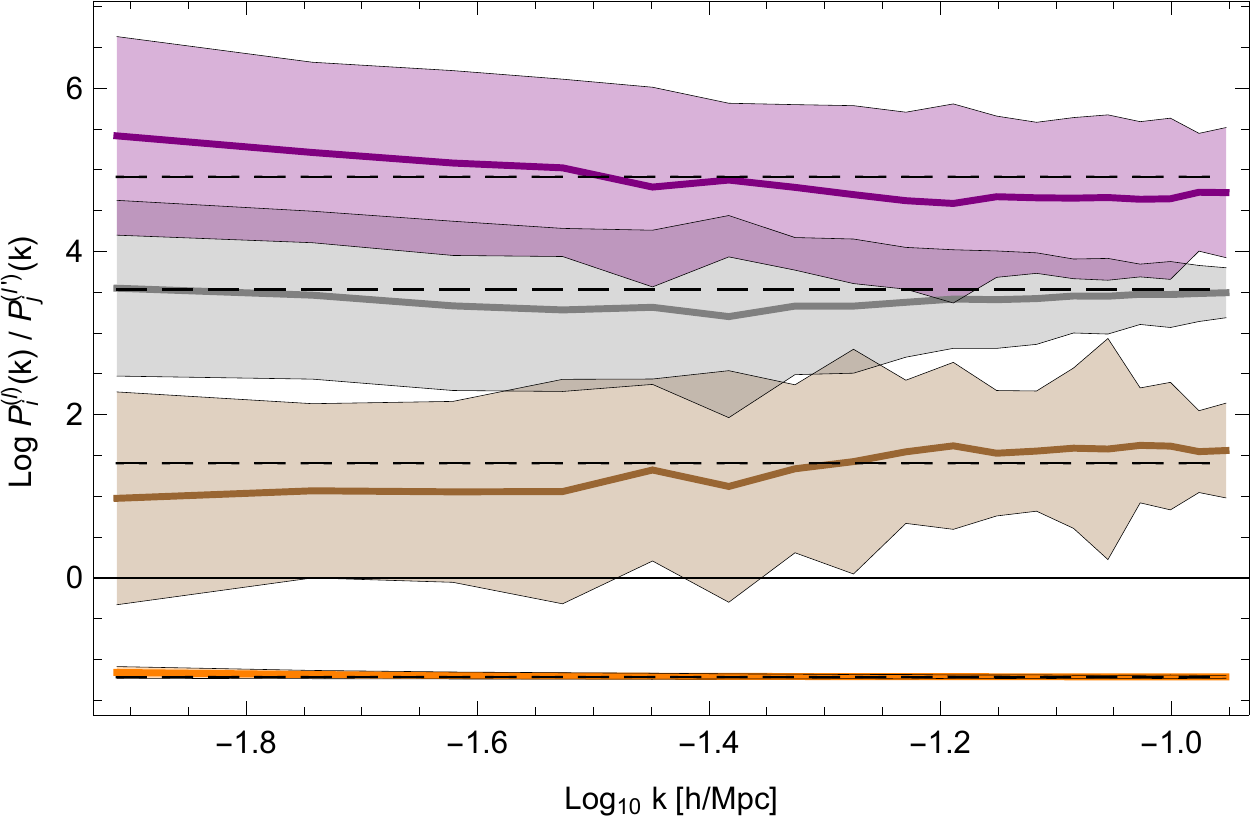}} 
\resizebox{7.5cm}{!}{\includegraphics{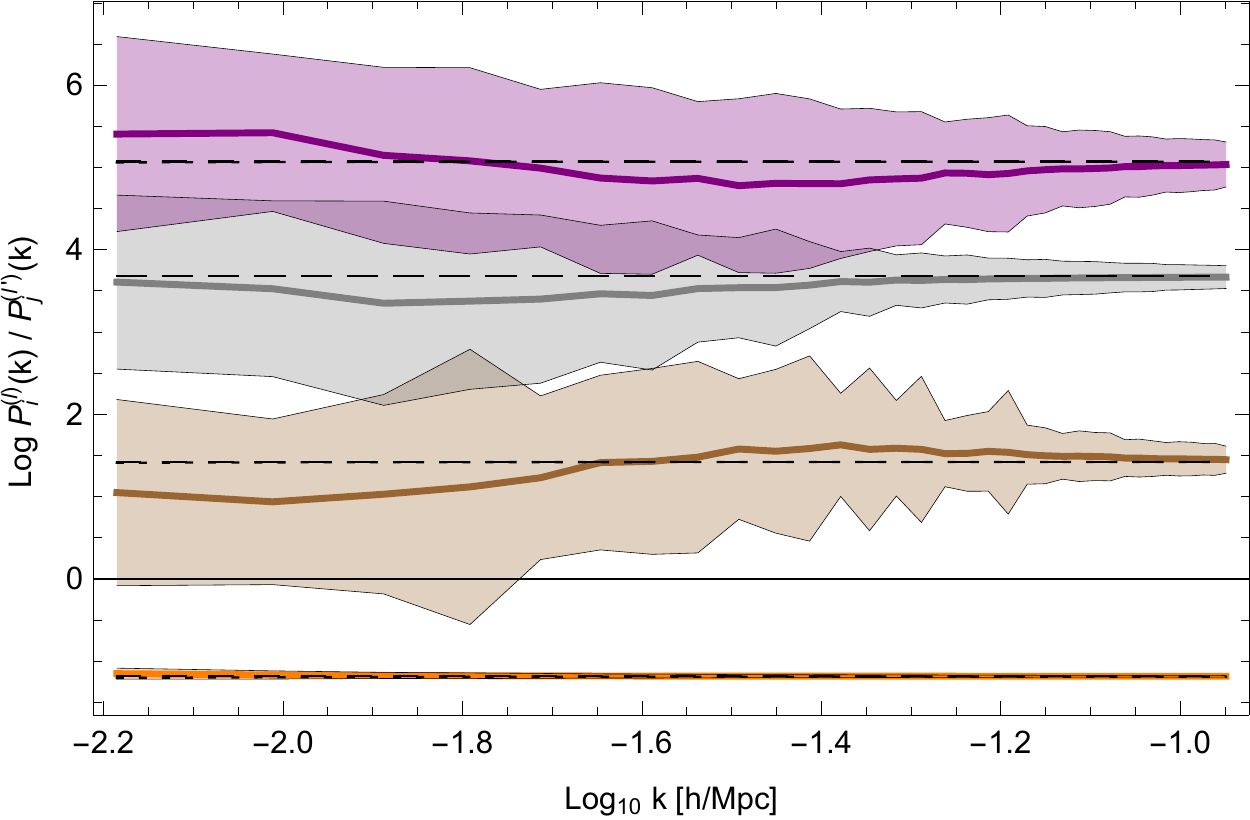}} 
\resizebox{7.5cm}{!}{\includegraphics{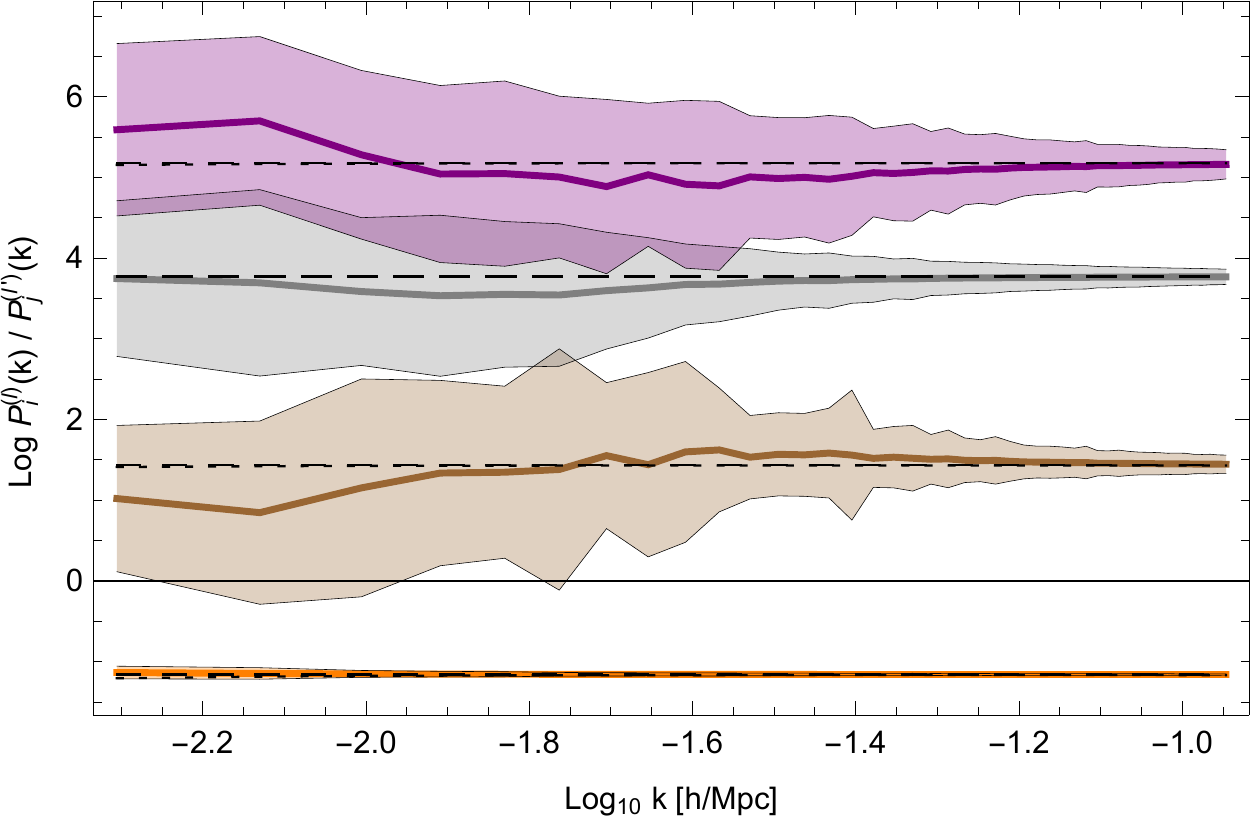}} 
\resizebox{7.5cm}{!}{\includegraphics{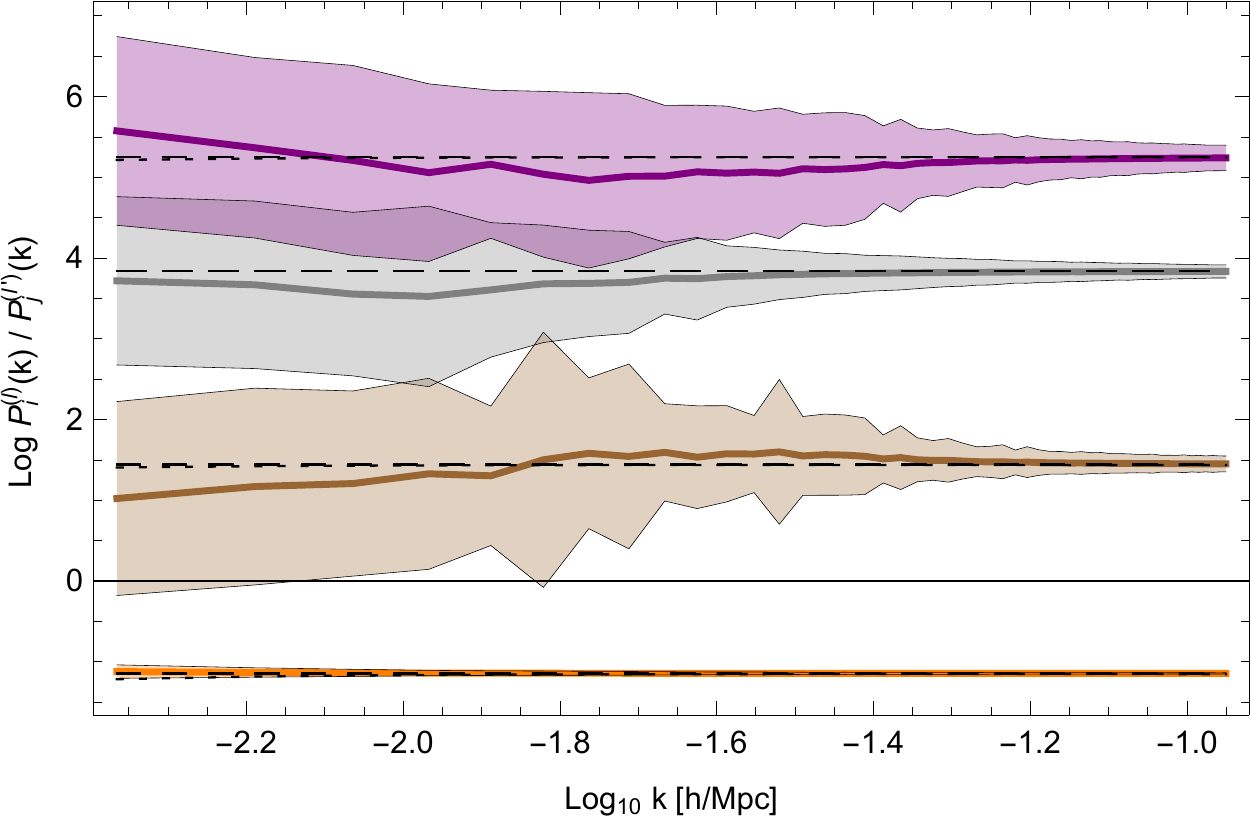}} 
\caption{
Sample means (thick lines) and variances (shaded regions) of $\log R_\mu$ for the four 
redshift slices -- see Eq. (\ref{Eq:ratios}).
From bottom to top, the lines denote the (log of the) ratios 
$R_1$ (orange in color version), $R_2$ (brown), $R_3$ (gray), and $R_4$ (purple), respectively.
The log ratios were displaced to improved visibility: $\log R_2$ by 2, $\log R_3$ by 4, and $\log R_4$ by 6. 
Dashed lines indicate the theoretically expected ratios of spectra 
without the Doppler term, and dotted lines denote a model without a Doppler, but with $f_{\rm NL}=10$.
}
\label{Fig:ratios}
\end{figure}

In the case of the simulated survey we consider in this paper, with two tracers, the situation is much simpler, 
and there is only one ratio of spectra that can be constructed (equivalent to $t_1$ defined above).
Moreover, in practice the anisotropic redshift-space power spectra reduce to only two degrees of freedom: 
the monopole and quadrupole (the $\ell \geq 4$ components are almost always too noisy and can be neglected).

Hence, with two tracers there are in total four possible ratios $P^{(\ell)}_i(k)/P^{(m)}_j (k)$, which 
we can organize in the following way:
\be
\label{Eq:ratios}
R_\mu (k) = \left\{ \frac{P^{(0)}_1}{P^{(0)}_2} \; , \; 
\frac{P^{(2)}_1}{P^{(2)}_2} \; , \; 
\frac{P^{(2)}_1}{P^{(0)}_1} \; , \; 
\frac{P^{(2)}_2}{P^{(0)}_2} \right\}
\ee
\begin{figure}
\resizebox{7.5cm}{!}{\includegraphics{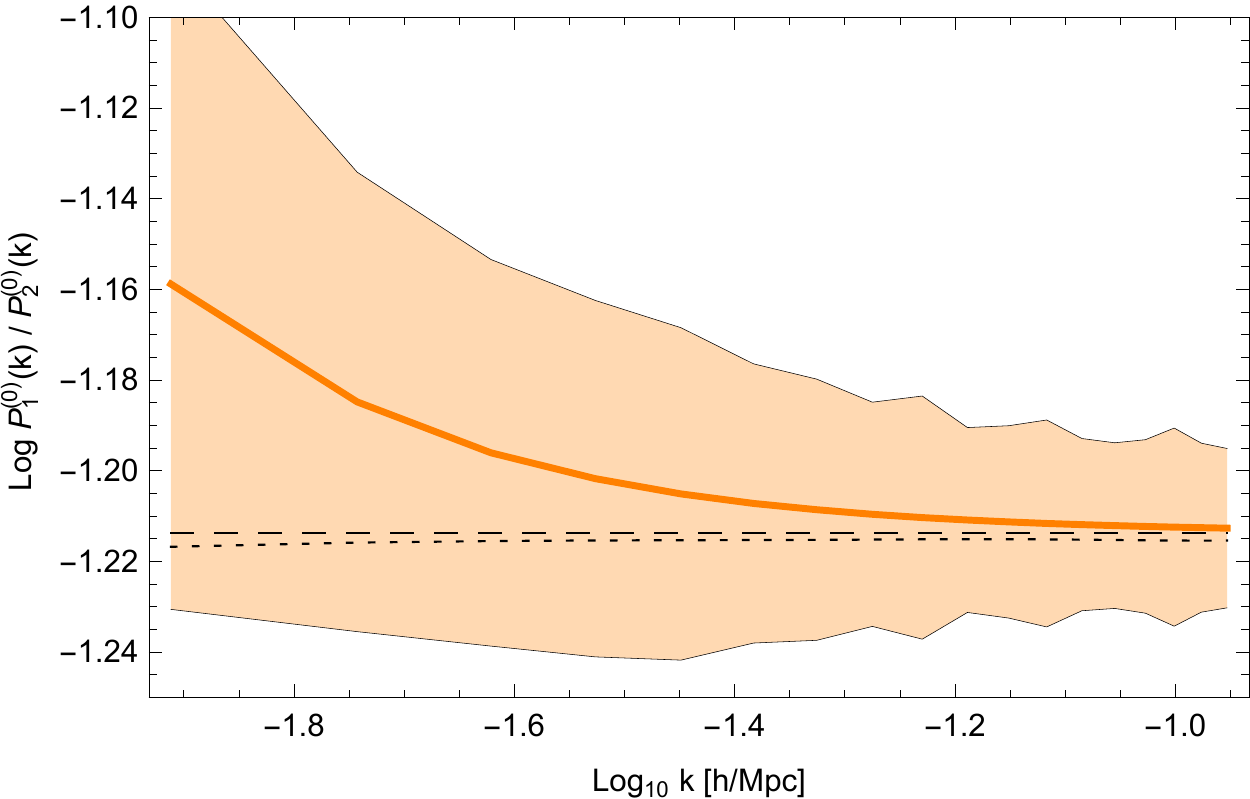}} 
\resizebox{7.5cm}{!}{\includegraphics{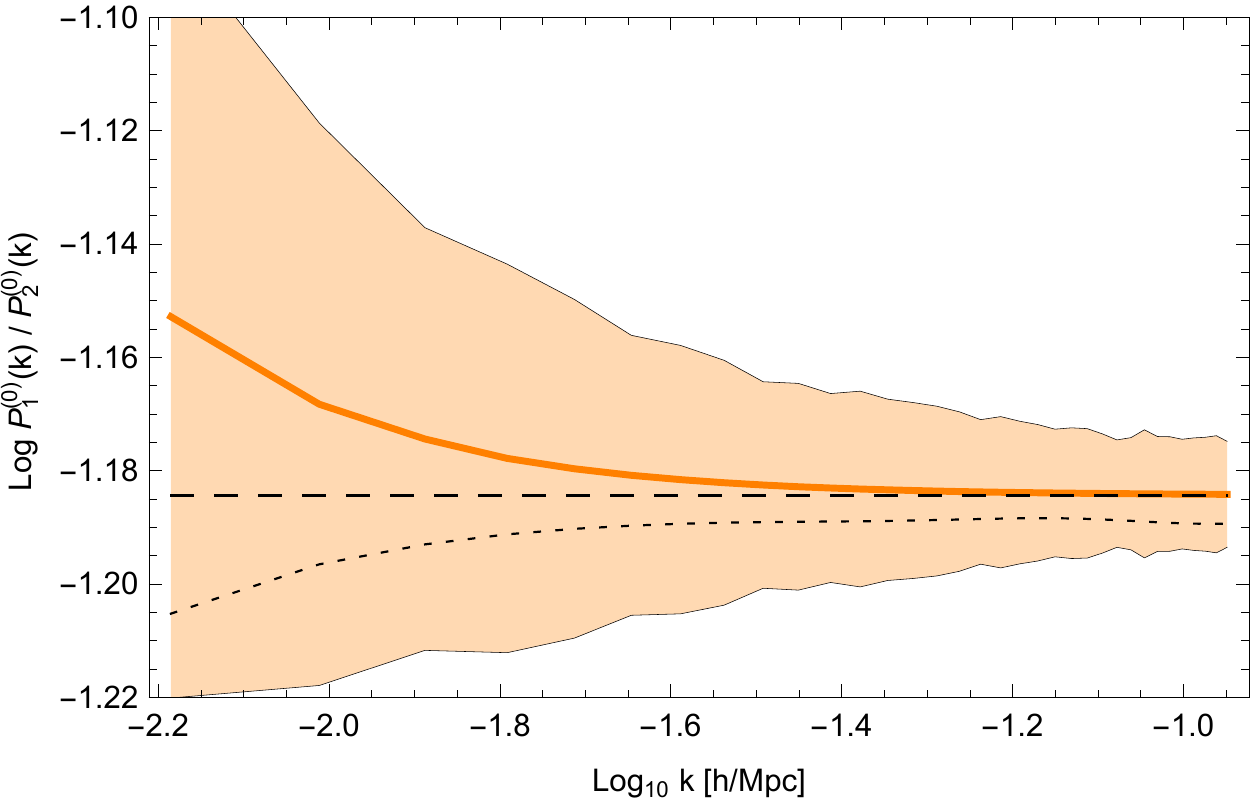}} 
\resizebox{7.5cm}{!}{\includegraphics{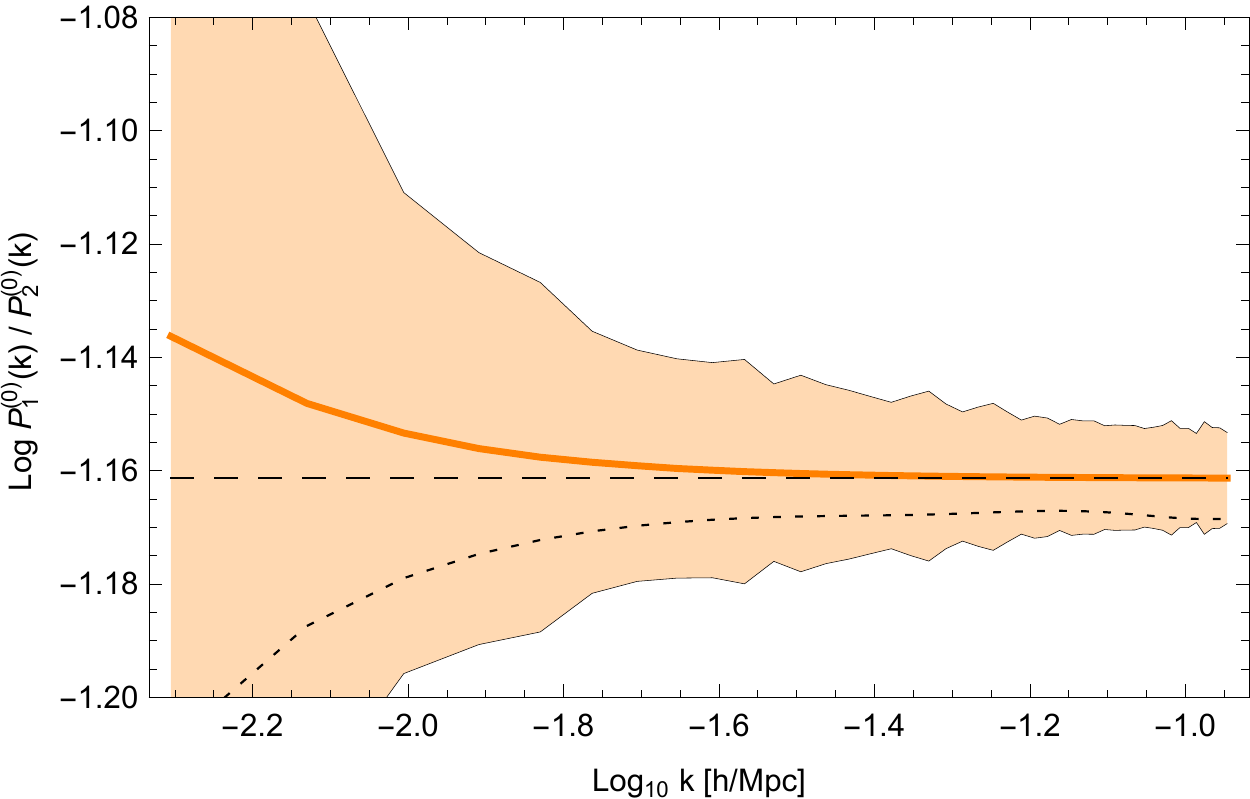}} 
\resizebox{7.5cm}{!}{\includegraphics{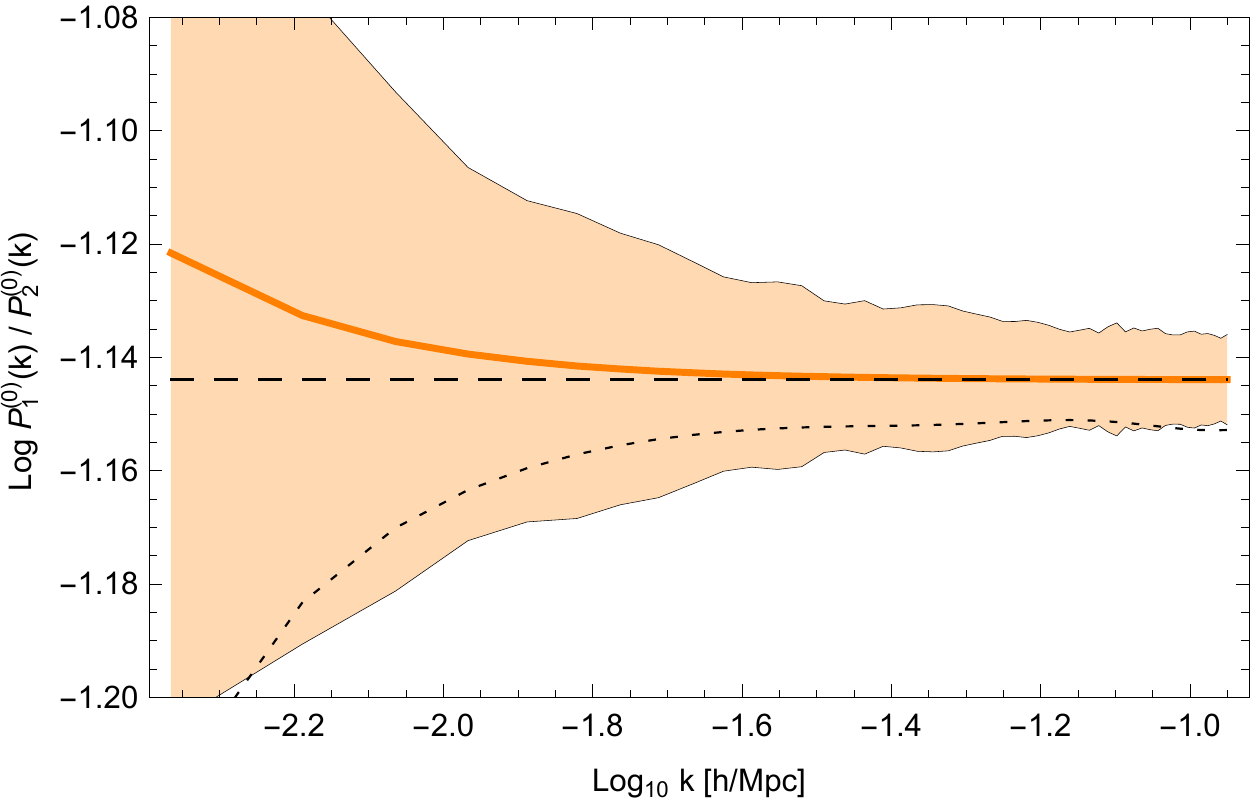}} 
\caption{
Sample means (thick lines) and variances (shaded regions) of $\log R_1$ (ratio of the monopoles of the
spectra of the two tracers), for the four redshift slices.
Dashed lines indicate the theoretically expected ratios of spectra 
without the Doppler term, and dotted lines denote a model without a Doppler, but with $f_{\rm NL}=10$.
}
\label{Fig:ratios1}
\end{figure}
In the Appendix we show the correlation matrix for these four ratios, obtained by means of our lognormal 
mock catalogs -- see Fig. \ref{Fig:grid}.

In Fig. \ref{Fig:ratios} we show the sample means and variances of $\log R_\mu$ for the four redshift slices (left panels).
From bottom to top on each of the left panels, the lines correspond, respectively, to the (log of) ratios 
$R_1$ (orange in color version), $R_2$ (brown), $R_3$ (gray), and $R_4$ (purple).
In order to improve the visualization we have displaced the $\log R_2$ by 2, $\log R_3$ by 4, and $\log R_4$ by 6. 
Moreover, since $\log R_1$ (ratios of the monopoles of the two tracers) has much lower variance compared with the other
ratios, we also show it separately, in Fig. \ref{Fig:ratios1}.

Notice, in particular, the dashed lines in Figs. \ref{Fig:ratios}-\ref{Fig:ratios1}, which denote the expected 
ratios without the Doppler, and the dotted lines, which denote the case without the Doppler term, but including pNG 
with an amplitude of $f_{\rm NL}=10$ . Using the ratios of monopoles of the two tracers (Fig. \ref{Fig:ratios1}) 
these features have an improved chance of detection, compared with either the 
direct measurements of the amplitudes of the spectra, or with
the ratios of the quadrupoles to the monopoles of each spectra -- upper lines in Fig. \ref{Fig:ratios}.

The fact that the ratio of monopoles increases with the inclusion of the Doppler term (dashed lines denote no Doppler term) 
is due to the fact that more highly biased tracers (in our scenario, the red galaxies) are relatively less sensitive to 
RSDs in general. In particular, if the two galaxies had the same bias, that ratio would be 1, with or without the 
Doppler term or pNG, meaning that the ratio of spectra carries no information in that case.
Notice also that the ratio of monopoles {\it decreases} when we include a positive value of $f_{\rm NL}$, 
due to the fact that pNGs are sensitive to $b-1$, which vanishes for the blue galaxies, but not for red galaxies.

\begin{figure}
\resizebox{10.cm}{!}{\includegraphics{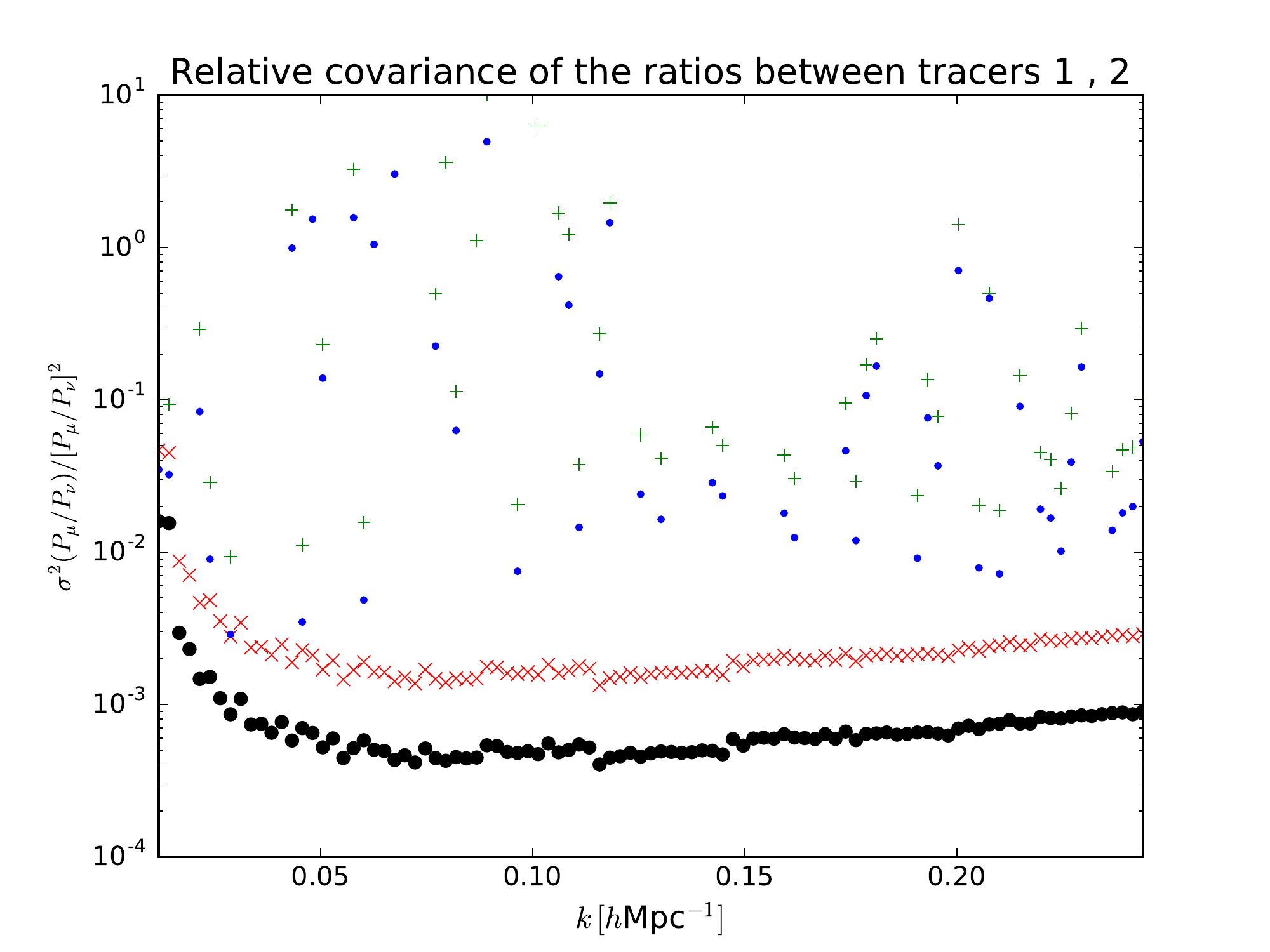}} 
\caption{Diagonals of the sample covariances of the ratio between the two monopoles (black dots and red crosses), 
and ratio of the two quadrupoles (blue dots and green symbols), for the redshift slice at $z=0.1$.
The dots correspond to the optimal estimation using the multi-tracer weights; $\times$ and $+$ symbols correspond
to the estimation using the FKP weights.
}
\label{Fig:FKP_MT}
\end{figure}

We should again stress that it is not sufficient to simply employ ratios of tracers, without proper weighting of the
different galaxy types and accounting for the fact that the two galaxy types share the same amount of cosmic variance. 
In that regard, the multi-tracer weights improve dramatically the measurements of 
ratios of spectra of different tracers, compared with the traditional (FKP) weights. 
In Fig. \ref{Fig:FKP_MT} we exemplify the noise suppression that follows from using
the multi-tracer weights, compared with the FKP weights, for the simulated redshift slice centered on 
$z=0.1$ (see Table II). The black solid dots denote the sample variances (diagonal part of the covariance)
for the ratios of monopoles of the two different tracers using the multi-tracer weights, 
while the red crosses show the same variance, but obtained using the FKP weights. 
Blue dots denote the variance of the ratios of the quadrupoles of the two tracers using the multi-tracer weights,
while the green $+$ symbols denote the same variance, but obtained using the FKP weights. 

\section{Local primordial non-Gaussianity}
\label{sec:PNG}
As evident from Eq.~\eqref{eq:pk_a}, Doppler effects are in general degenerate with any scale-dependence that could arise from modifications of gravity, neutrino effects, or non-Gaussianity.
In fact, there is a particular scale-dependent bias correction on large scales which is induced by primordial non-Gaussianity (pNG) 
through the squeezed limit of the bispectrum. These non-Gaussianity distort halo bias, and that in turn implies in a scale-dependent galaxy bias (see e.g. ~\cite{Matarrese:2000iz, Dalal:2007cu, Matarrese:2008nc, Slosar:2008hx, Desjacques:2016bnm}), i.e.
\be
\label{b_fNL}
b_i \to  b_i+\Delta b_i^{nG} (k) \quad \quad {\rm where}  \quad \quad \Delta b_i^{nG} (k) = 3 \, f_{\rm NL} \, (b_i-1) \frac{\Omega_m \, H_0^2 \, \delta_c }{c^2 \, k^2 \, T(k) \, D(z) } \; .
\ee
Here $f_{\rm NL}$ is the local-type amplitude of non-Gaussianity, $T(k)$ is the
transfer function (normalized to unity on large scales), and $D(z)$ is the growth function normalized to $D(0)=1$ 
(we take the critical density $\delta_c=1.7$).
Since the transfer function is nearly constant on large scales  ($k \le k_{\rm eq} \ll 0.1 \, h \, {\rm Mpc}^{-1}$), in that limit
the pNG bias correction and the Doppler one both scale in the same way, as $\sim k^{-2}$. However, whereas the
redshift dependence of the Doppler term scales as $\sim f(z)/r(z)$, the pNG bias scales as $\sim 1/D(z)$, 
which means that measurements taken at different redshifts can distinguish between the two effects.

A few key physical parameters can be probed through galaxy surveys only on the largest observable scales: primordial non-Gaussianity and relativistic effects, in particular, are only manifested through features on scales of the order of the Hubble radius.
Measurements of the CMB already constrained the non-Gaussianity parameter $f_{\rm NL}$ to be 
$\lesssim 10$~\cite{PlanckNG}, and future large-scale galaxy surveys aims to measure it with a sub-unity 
precision~\cite{Alvarez:2014, Raccanelli:2014fNL,Raccanelli:2015fNLISW,dePutter:2014} 
(see also \cite{Camera:2014bwa, Camera:2014sba} where local and non-local projection corrections are taken into account).
Measuring such percent-level distortions on galaxy power spectra, on top of cosmic variance and systematics, 
is a major challenge. One of the ways to reach such target is to make use of the multi-tracer technique~\cite{Ferramacho:2014} introduced in Section~\ref{sec:data}.

The additional contribution to the RSD operator introduced in Section~\ref{sec:alpha} induces large-scale 
modifications to the galaxy power spectrum (through the $k^{-2}$ dependence), so it will be somewhat 
degenerate with local pNG effects. In Fig. \ref{Fig:ratios1} the dotted lines denote the theoretical expectation for the ratio of the
monopoles of the two types of galaxies, at the four redshift slices, for the case $f_{\rm NL}=10$. The lower left plot, in
particular, shows that for the third redshift slice (centered on $z=0.5$) the Doppler effect has a signature 
approximately opposite to that of pNG -- in fact, an $f_{\rm NL}\simeq -10$ would produce a signature not unlike 
that of the Doppler dipole term in the case under study, at that redshift.
Hence, it makes sense to investigate the impact of these corrections 
on parameter estimation; we focus on measurements of the pNG parameter $f_{\rm NL}$ in the local, 
or squeezed, limit.

While a careful analysis including the $\alpha$ term with its angular and redshift dependence can 
disentangle the two effects, a naive analysis (neglecting the velocity terms) could mistake Doppler effects 
for pNG (or vice-versa). Therefore, when computing our results, after investigating the detectability of 
such effects, we will include the $f_{\rm NL}$ parameter in our analysis, and study how its constraints 
change due to the inclusion or not of Doppler effects.

\section{Fisher matrix}
\label{sec:FM}
We would now like to use all the information in our mock survey to extract constraints on the Doppler terms,
on pNGs, as well as other parameters with whom they can be degenerate. In addition to the 
ratios of quadrupoles to monopoles, $P^{(2)}_i/P^{(0)}_i$, proposed by 
Percival \& White, we will also employ the ratios of the monopoles and quadrupoles of different tracers -- see Fig. \ref{Fig:ratios}. 
The correlation between these four ratios is shown in the Appendix (see Fig. \ref{Fig:grid}).

In this Section we will compute the Fisher matrix for a set of parameters, and derive
the constraints for those parameters by inverting that Fisher matrix.
The procedure can be summarized as follows: first, the data covariance is effectively inverted, 
generating the data Fisher matrix; next we project the data Fisher matrix into the Fisher matrix for the parameters, 
using the derivatives of the multipoles of the power spectra with respect to the parameters (these derivatives are
computed analytically); finally, the parameter-space Fisher matrix is combined with priors, and that matrix is 
inverted to produce the covariance matrix for the parameters.

In order to show how the different ratios contribute to the constraints, we show results 
corresponding to the use of three combinations of these observables: 
\begin{itemize}
\item (i) the ratios $R_3$ and $R_4$ (``PW'');
\item (ii) the multi-tracer ratios $R_1$ and $R_2$ (``MT'');
\item (iii) the four ratios above (``All'').
\end{itemize}
In each case, the data (sample) covariance matrix is inverted to obtain the data Fisher matrix,
$F^a (\mu, k ; \nu , k') = \{ {\rm Cov}[ R_\mu(k) , R_\nu(k')] \}^{-1}$, with $a=$ {\it PW} ($\mu,\nu = \{3,4\}$), {\it MT} 
($\mu,\nu = \{1,2\}$), or {\it All} ($\mu,\nu = \{1,2,3,4\}$).
With $n_k$ Fourier bins in a given redshift slice,
the covariance and data Fisher matrix has dimensions $N \times N$, 
where $N = 2 n_k$ in the PW and MT cases, and $N = 4 n_k$ when all four ratios are considered.
In order to keep the contamination from non-linear scales limited to a minimum, we cut-off the information
above $k=0.1 \, h$ Mpc$^{-1}$ on all redshift slices.

Once the data Fisher matrices have been computed, they can be projected into Fisher matrices for the parameters, $\theta^i$.
We have considered the following set of parameters: the biases ($b_i$), the amplitudes of the Doppler terms ($\alpha_i$), the matter growth rate ($f$), and the amplitude of pNGs ($f_{\rm NL}$).
We model the matter growth rate, as a function is redshift, in terms of the parametrization
$f(z) = \Omega_m(z)^\gamma$, where $\gamma$ is a free parameter ($\gamma \simeq 0.55$ for $\Lambda$CDM).

The most optimistic scenario, the biases are a single parameter, for all redshift slices, and the matter growth rate is also
determined by a single parameter ($\gamma$). In this case, there are six parameters in the Fisher matrix,
given by $\theta^i = \{ b_1, b_2, \alpha_1, \alpha_2, \gamma, f_{\rm NL} \} $.
However, a more realistic approach is to take the biases of the two galaxy types to be unknown on each 
redshift slice, in which case we would have 12 parameters: 
$\theta^i = \{ b_1(z), b_2(z), \alpha_1, \alpha_2, \gamma, f_{\rm NL} \} $.
Finally, an even more conservative approach is to take as free parameters not only the biases, but the matter growh rate on each 
redshift slice, in which case we would have 15 parameters: 
$\theta^i = \{ b_1(z), b_2(z), \alpha_1, \alpha_2, \gamma (z), f_{\rm NL} \} $.

The final Fisher matrices are then given by:
\be
\label{Eq:Fish}
F_{i j} = \sum_{\mu,k} \sum_{\nu,k'} 
\frac{\partial R_\mu (k)}{\partial \theta^i} F (\mu, k ; \nu , k') \frac{\partial R_\nu (k')}{\partial \theta^j} \; ,
\ee
We compute the Fisher matrices on each individual slice, and add the information from different slices 
in order to obtain constraints for the survey volume as a whole.

It is important to notice that, since our observables are ratios of spectra, 
there is a basic degeneracy in the space of parameters: one can
scale the parameters by a constant, and the ratios will be invariant under 
such a transformation. The only variable which breaks this degeneracy is the matter growth rate, which we
have assumed given by the fit in terms of $\gamma$.
However, our resulting Fisher matrices are still very nearly singular due to this basic degeneracy.
This can be fixed by adding any extra information (a prior) that breaks the scale invariance
-- e.g., an interval for the bias, or theoretical limits for $\alpha$.
For this reason we have adopted weak priors on the biases ($\sigma_{b_1}^\Pi = \sigma_{b_2}^\Pi = 1$), on the
Doppler terms ($\sigma_{\alpha_1}^\Pi = 8, \sigma_{\alpha_2}^\Pi = 4$), on the matter growth rate power index 
($\sigma_\gamma^\Pi = 1$) and on the 
amplitude of pNGs ($\sigma_{f_{\rm NL}}^\Pi = 10$), through a prior 
$\Pi = {\rm diag} [(\sigma_{b_1}^\Pi)^{-2},(\sigma_{b_2}^\Pi)^{-2},(\sigma_{\alpha_1}^\Pi)^{-2},(\sigma_{\alpha_2}^\Pi)^{-2},(\sigma_\gamma^\Pi)^{-2},(\sigma_{f_{\rm NL}}^\Pi)^{-2} ]$.

Table II shows the uncertainties in the six parameters which result from different combinations
of the information contained in the ratios of tracers. The uncertainty on any given parameter is, as usual, obtained by 
marginalizing over every other variable -- i.e., by inverting the Fisher information, 
$\sigma^2(\theta^i) = (F ^{-1})_{i i}$.

\begin{table}
\begin{center}
\begin{tabular}{|c|c|cccccc|}
\hline
$ $ & Fiducial values & $b_1=1.0$ & $b_2=2.0$ & $ \alpha_1 = 4.0$ & $\alpha_2 = 2.0$ & $\gamma = 0.55$ & $f_{\rm NL} =0.0$ \\
\hline
\hline
Case/row & Fisher Information & $\sigma(b_1)$ & $\sigma(b_2)$ \; & \;  $ \sigma(\alpha_1)$ & $\sigma(\alpha_2)$ \; & \; $\sigma(\gamma)$ & $\sigma(f_{\rm NL})$ \\
\hline
(1) & $F^{PW}$ & 0.069 & 0.17 &  9.7 &  25.9 & 0.13 & 138 \\
(2) & $F^{MT}$ & 0.83 &  1.37 &  8.3 &  40.3 &  0.58 & 7.0\\
(3) & $F^{All}$ & 0.027 & 0.055 & 3.9 &  22.9 &  0.05 & 5.8 \\
\hline
$ $ & Prior ($\Pi$) & 1.0 & 1.0 & 8.0 &  4.0 & 1.0 & 10.0 \\
\hline
(4) & $F^{PW}$ + $\Pi$  & 0.065 & 0.15 &  5.84 &  3.94 & 0.12 & 9.97 \\
(5) & $F^{MT}$ + $\Pi$ & 0.41 &  0.67 &  2.2 &  3.97 &  0.28 & 5.6 \\
(6) & $F^{All}$ + $\Pi$ & 0.025 & 0.051 & 1.21 &  3.93 &  0.046 & 5.0 \\
\hline
(7) & $F^{All}$ + $\Pi$ + fixed ($\gamma$, $f_{\rm NL}$)  & 0.013 & 0.022 & 1.05 & 3.93 &  -- & -- \\
(8) & $F^{All}$ + $\Pi$ + fixed ($b_1$, $b_2$) & -- & -- &  1.08 &  3.93 & 0.016 & 2.53 \\
(9) & $F^{All}$ + $\Pi$ + fixed ($b_1$, $\alpha_1$) & -- & 0.0046 & -- &   3.29 & 0.022 & 3.7 \\
(10) & $F^{All}$ + $\Pi$ + fixed ($b_2$, $\alpha_2$) & 0.0024 & -- & 0.94 & -- & 0.020 & 3.8 \\
\hline
(11) & $F^{All}_{z_1}$ + $\Pi$ 				& 0.41 & 0.82 &  1.61 &  3.95 & 0.41 & 10.0 \\
(12) & $F^{All}_{z_1+z_2}$ + $\Pi$ 			& 0.12 & 0.25 &  1.29 &  3.94 & 0.15 & 9.4 \\
(13) & $F^{All}_{z_1+z_2+z_3}$ + $\Pi$ 		& 0.049 & 0.098 &  1.24 &  3.93 & 0.075 & 7.2 \\
(14) & $F^{All}_{z_1+z_2+z_3+z_4}$ + $\Pi$ 
= $F^{All}$ + $\Pi$ 					& 0.025 & 0.051 & 1.21 &  3.93 &  0.046 & 5.0 \\
\hline
\hline
(15) & $F^{All}$ + $\Pi$ [$b_i(z)$ ] & -- & -- & 1.31 &  3.86 &  0.92 & 5.1 \\
(16) & $F^{All}$ + $\Pi$ [$b_i(z)$, $\gamma(z)]$ & -- & -- & 1.76 &  3.86 &  -- & 6.0 \\
\hline
\end{tabular}
\caption{Uncertainties in the parameters when different combinations of the Fisher information 
and priors are taken into account.
We only include the information from modes up to $k=0.1 \, h \, {\rm Mpc}^{-1}$ in order to limit ourselves to the
linear regime. Rows 1-14 correspond to the case when we take galaxy bias and the matter growth rate index 
$\gamma$ to be free parameters, but with a single value for all redshift slices.
On row 15 we allow the biases to assume values on each redshift slice, and on row (16) we assume, in addition, 
that the value of $\gamma$ is independent on each slice.
As one can see from rows 11-14 of the table, most of the information about the Doppler 
is in fact coming from the first redshift slice -- although the higher-redshift slices help to determine bias and
the matter growth rate, which helps breaking degeneracies. 
It is also interesting to notice that, although the multi-tracer ratios (MT) of monopoles are much better at detecting
the Doppler, the direct comparison of the quadrupole with the monopole (PW) for individual tracers performs
better at detecting the matter growth rate, expressed through the parameter $\gamma$. 
This is because the monopole inherits contributions from both the
bias and from matter growth rate $f$, while the amplitude of the quadrupole is directly proportional to $f$.
}
\end{center}
\label{tab:resFM}
\end{table}

\begin{figure}
\resizebox{8.0cm}{!}{\includegraphics{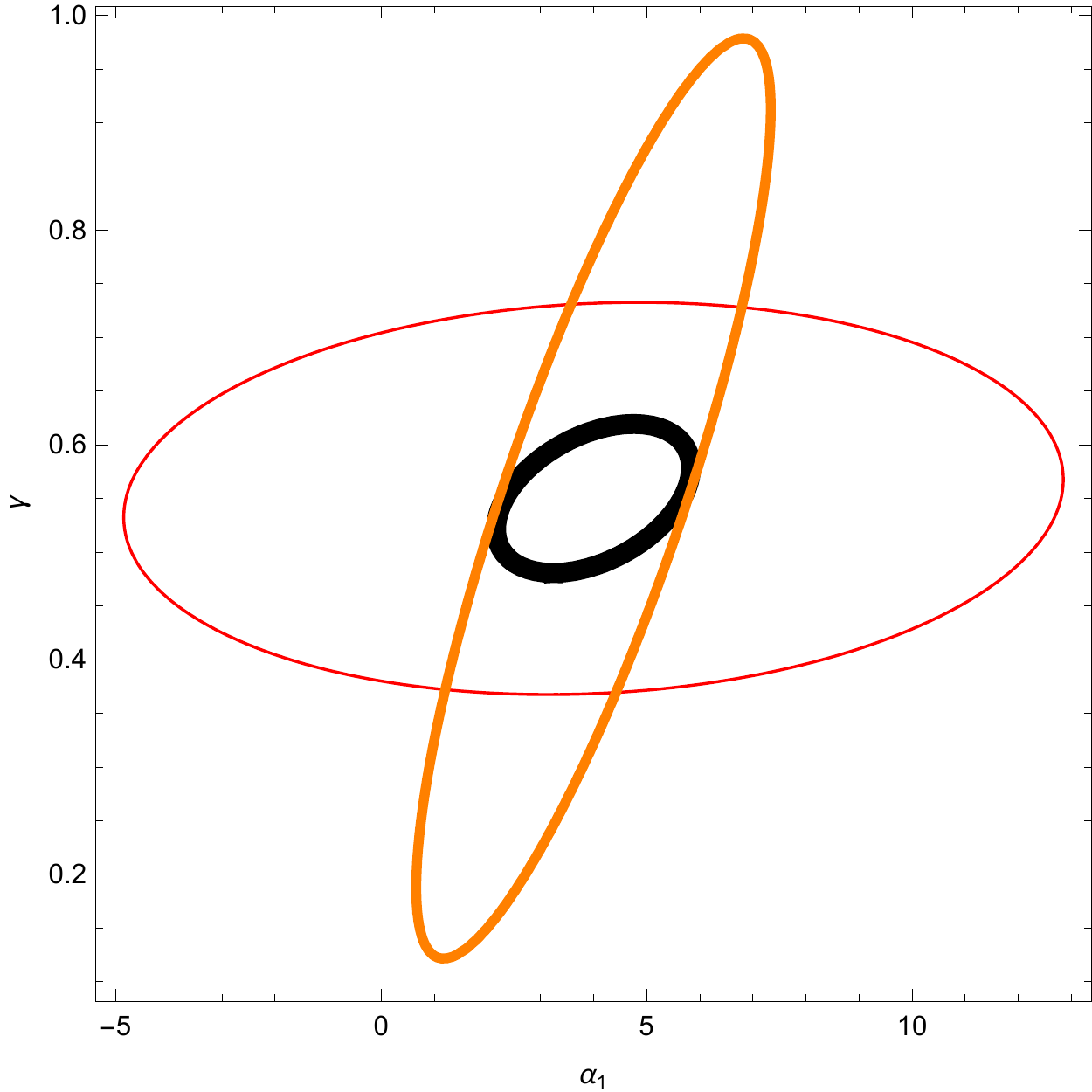}}
\caption{Red (thin) line: 1-$\sigma$ constraints using the ratios of quadrupoles and monopoles 
(the $PW$ observables), together with the prior defined in Table II. 
Orange line: constraints obtained with the ratios of monopoles of the two tracers (the $MT$ observables), plus 
the prior. Black (thick) line: $PW + MT$ ($All$), with the prior. Here, bias and the matter growth index $\gamma$ are
assumed to be constant on all slices.}
\label{Fig:ag}
\end{figure}

In our simulations, the ratios of the quadrupoles and monopoles of the same tracers (the {\it PW} observables) 
perform slightly better at constraining galaxy bias and the matter growth rate, compared the ratios of spectra 
of different tracers (the {\it MT} observables). The multi-tracer information, on the other hand,
is more powerful to constrain the Doppler and $f_{\rm NL}$ terms. In particular, as shown in 
Fig. \ref{Fig:ag}, the $PW$ ratios are better at measuring the matter growth rate index $\gamma$, while the 
$MT$ ratios are more sensitive to the Doppler term.
When the two types of observables are combined ($All$), the strengths of both approaches help break the internal
degeneracies, and we obtain a $\sim 3.3 \, \sigma$ detection of $\alpha_1$ in the optimist case (row 6), a
$\sim 3.0 \, \sigma$ detection in the conservative case (row 15), and a
$\sim 2.3 \, \sigma$ limit in the ultra-conservative case (row 16). 
The constraints on $f_{\rm NL}$, on the other hand, seem less sensitive to our assumptions about bias and the 
matter growth rate compared with the constraints for the Doppler term.
Notice that there is no scenario in which 
there are hints of $\alpha_2$: this is partly due to the higher bias of the second tracer species 
(red galaxies, $b_2=2.0$), which makes the relative effect of the Doppler weaker for that tracer.

The remaining entries of Table II 
show several cases where we fix some parameters, and marginalize against others.
By comparing the rows 1-3 with rows 5-7, we are able to see how the constraints are affected by the prior: 
in particular, it can be seen that even a weak prior on $\alpha_2$ 
helps pin down both $\alpha_1$ and $f_{\rm NL}$, especially when all the ratios of spectra are taken into account.

\begin{figure}
\resizebox{7.5cm}{!}{\includegraphics{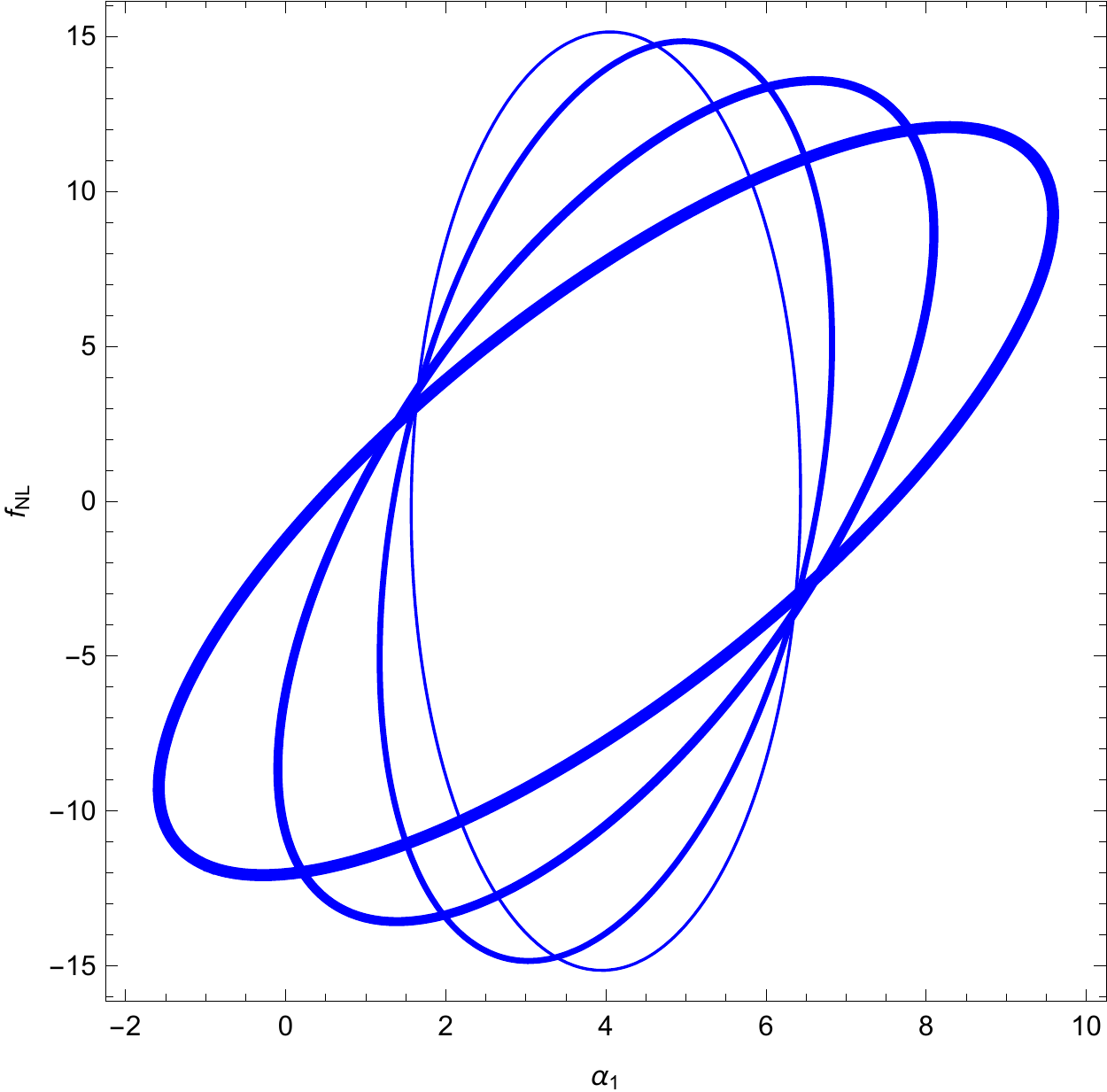}}
\resizebox{7.5cm}{!}{\includegraphics{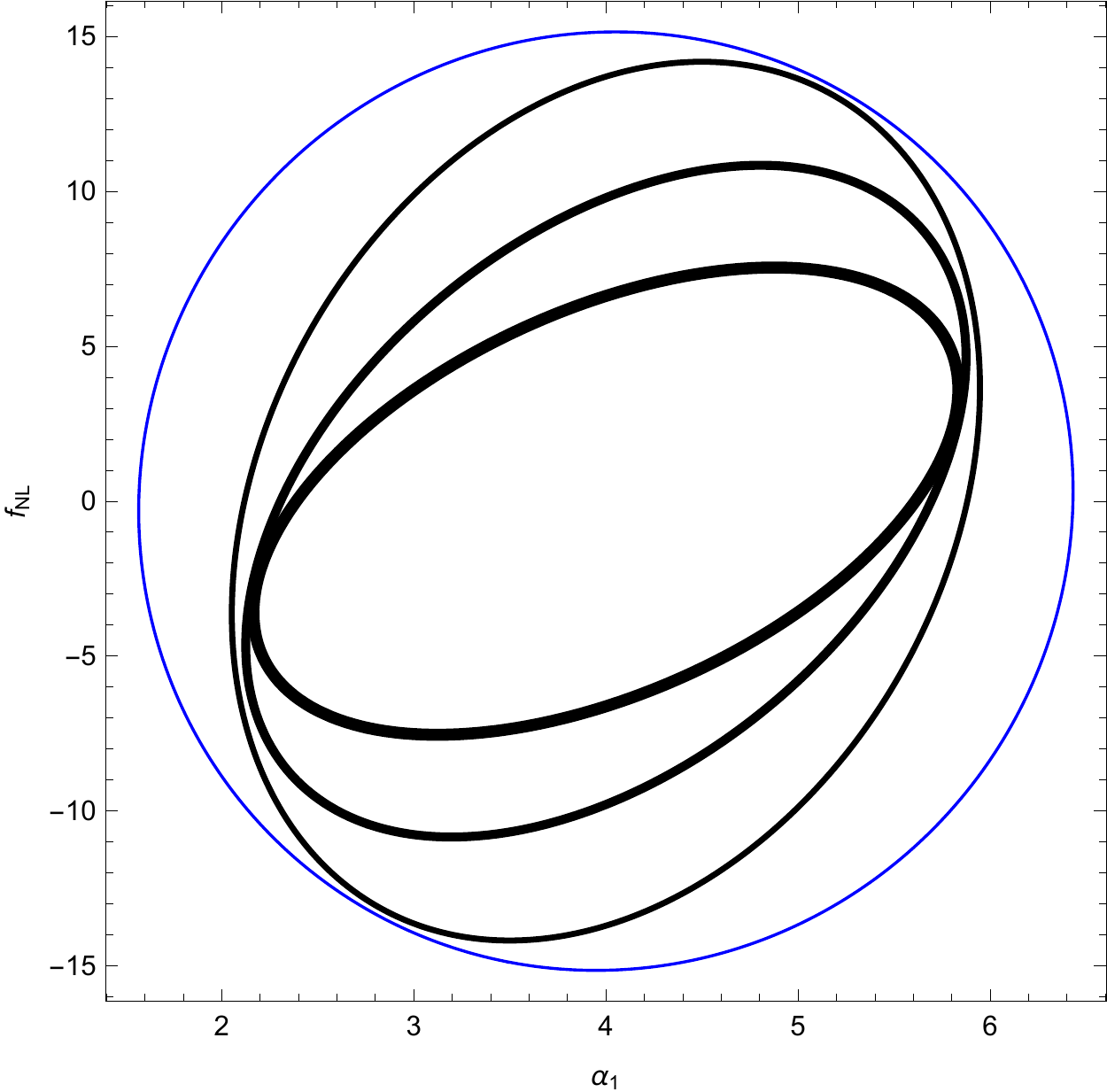}}
\caption{Left: 1-$\sigma$ constraints using all the ratios ($PW + MT$), together with the prior defined in Table II, for the
four redshift slices (from $z_1=0.1$, thinnest, to $z_4=0.7$, thickest line).
Right: cumulative constraints as we combine the successive redshift slices. The thin blue line on both plots 
correspond to the constraint from the first redshift slice (plus the prior).}
\label{Fig:afnl}
\end{figure}

Rows 11-14 of Table II show the cumulative information as we add the redshift slices. Although most of 
the information about the Doppler term comes from the first redshift slice (for which the observables are most 
sensitive to $\alpha$ due to the $1/r$ scaling), the other slices play an important role in breaking degeneracies 
with the bias and with the matter growth rate. The way in which the different redshift slices are sensitive to each parameter 
is shown in Fig. \ref{Fig:afnl}, where on the left panel we show how each slice constrains $\alpha_1$ and 
$f_{\rm NL}$, and on the right panel we show the cumulative effect of combining the information from the slices.

\section{Discussion and conclusions}
\label{sec:conclusions}

We have investigated the detectability of the velocity terms in galaxy clustering, as well as their degeneracy with primordial
non-Gaussianity.
The velocity effects are the combination of local peculiar velocity, mode-coupling, Doppler lensing and 
lightcone volume effects, which are often neglected in galaxy clustering analyses.
Although present in the original~\cite{Kaiser:1987} formulation of RSD, as well as the 
review~\cite{Hamilton:1998}, their influence were first investigated in~\cite{Szalay:1997, Papai:2008, Raccanelli:2010}, 
and more recently~\cite{Jeong:2011as, Raccanelli:2016doppler}. Here we propose to measure these
two effects using a mock survey of two galaxy types (blue galaxies, with $\alpha_1=4.0$ and bias $b_1=1.0$,
and red galaxies, with $\alpha_2=2.0$ and bias $b_2=2.0$).

It was pointed out by~\cite{Raccanelli:2016doppler} that these effects induce a $k^{-2}$ scale-dependence 
in the galaxy power spectrum, making them potentially degenerate with primordial non-Gaussianity 
(see also \cite{Jeong:2011as}). Fig. \ref{Fig:afnl} shows the nature of the degeneracy between the 
Doppler effects (parametrized in that case by $\alpha_1$) and $f_{\rm NL}$. The direction of this degeneracy
means that a small increase in $\alpha_1$ can be partially compensated by a small increment in $f_{\rm NL}$. 
This is in fact already visible from the right panels of Fig. \ref{Fig:ratios}, which show the ratio $P^{(0)}_1(k)/P^{(0)}_2(k)$.
The fact that this ratio increases with respect to the case of no Doppler term (dashed lines) is due to the fact
that more highly-biased tracers (in our scenario, the red galaxies, type=2) are relatively less sensitive to RSDs in general.
On the other hand, this ratio {\it decreases} when we include a positive value of $f_{\rm NL}$ 
(in the example shown in those plots, $f_{\rm NL}=10$), due to the fact that pNGs are sensitive to $b-1$, 
which vanishes for the blue galaxies ($b_1=1$), but not for red galaxies ($b_2=2$).

Despite this degeneracy, the fact that the Doppler and pNG effects evolve differently with redshift 
means that we can use galaxy maps at several redshifts to help break those degeneracies.
Fig. \ref{Fig:afnl} shows that the combining low-redshift surveys with intermediate- or high-redshift galaxy 
surveys will help to disentangle the Doppler effect from pNGs.

Our final results indicate that, in principle, a $\sim 3 \sigma$ measurement of a value of $\alpha=4$ 
is possible from a very large-area, low- and intermediate-redshift galaxy survey that distinguishes betwen 
different tracers of large-scale structure. If, on the other hand, one employs a galaxy sample for 
which $\alpha=3$, that measurement would be at $\sim 2 \sigma$.
We also showed that it is possible to constrain, at the same time as the Doppler term, local-type pNG, by up 
to $|f_{\rm NL}| \lesssim 5$.
The future prospect of very complete photometric samples over large areas means it should be possible 
to select sub-samples of galaxies in such a way that the Doppler term is enhanced, thus helping 
confirm the signal of those effects on galaxy clustering and limiting the scope of degeneracies with pNG.

In conclusion, we have shown that with very large-area surveys the Doppler terms are 
in principle detectable and can be distinguished from local-type primordial non-Gaussianities, 
and should therefore be included in future forecasts for precise measurements of galaxy clustering.

\vspace{0.3 cm}

{\bf Acknowledgments:}\\
We would like to thank Alvise Raccanelli for many useful discussions and a critical reading of the drafts. 
We also thank Donghui Jeong and Licia Verde for comments.
We both grateful for the Galileo Galilei Institute for Theoretical Physics (GGI), where this project 
was initiated, for its hospitality.
LRA thanks FAPESP (grant 2015/17199-0) and CNPq for financial support.
During the preparation of this work DB was supported by the Deutsche 
Forschungsgemeinschaft through the Transregio 33, The Dark Universe. 

\section*{Appendix: correlations and covariances}

\begin{figure}
\resizebox{12.0cm}{!}{\includegraphics{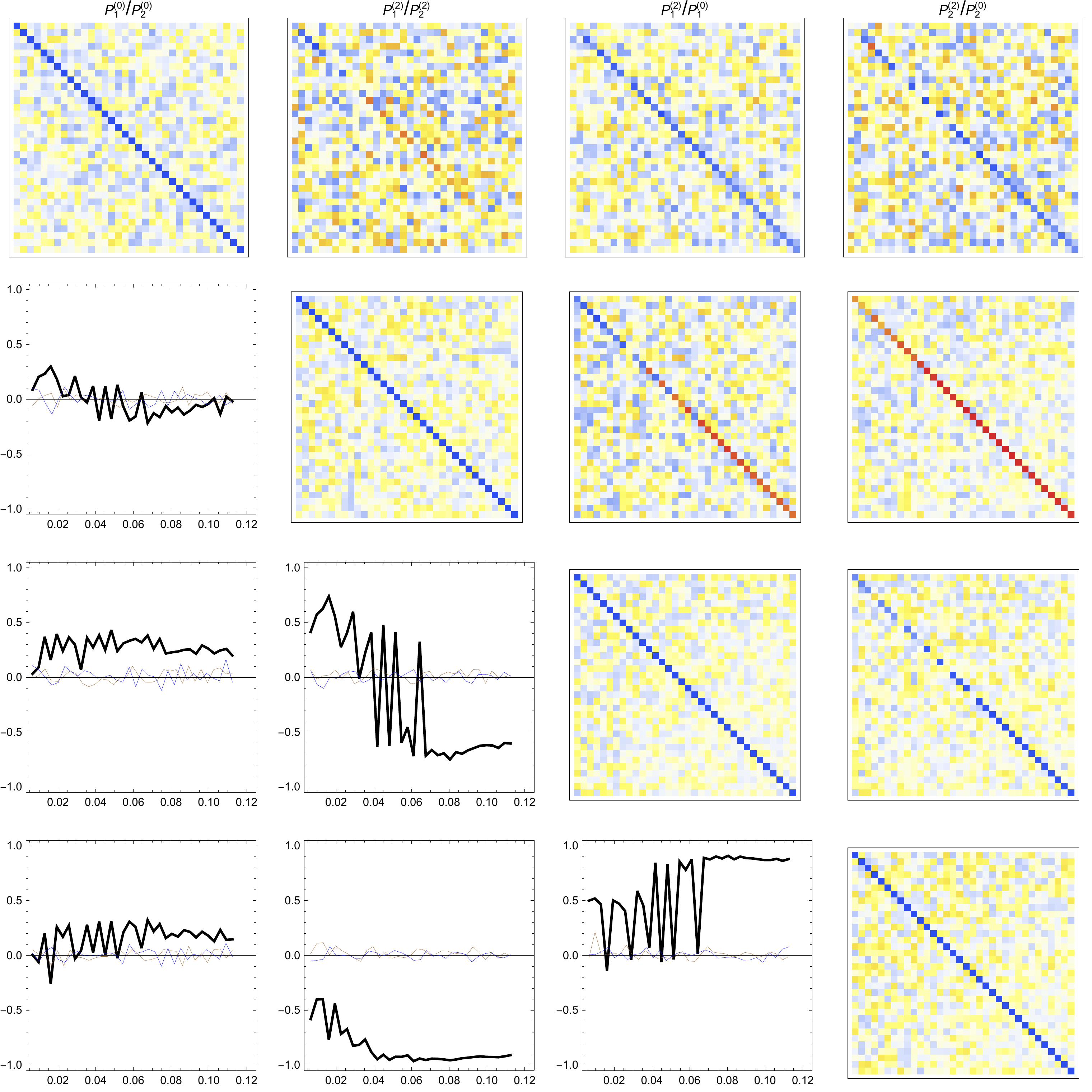}}
\caption{Correlations between the four possible ratios of monopoles and quadrupoles of the two tracers 
[see Eq. (\ref{Eq:ratios})], obtained for the
300 mocks corresponding to the second redshift slice (at $z=0.3$).
The upper triangle of the table shows the correlation matrices, and the lower triangle shows
the diagonal (thick, black in color version) and two first sub-diagonals (red and blue lines) of the correlation matrices.
The diagonal of the auto-correlations are not shown because the off-diagonal terms are negligible.
It is visible that the first ratio is almost entirely uncorrelated with the others, while the other three ratios are
slightly correlated or anti-correlated with each other, especially on small scales.}
\label{Fig:grid}
\end{figure}

In Figure~\ref{Fig:grid} we show the correlations between the four ratios used as observables. 
The upper triangle shows the actual correlation matrices, and the lower triangles show the 
diagonal of the cross-correlation matrices, as well as the second and third sub-diagonals 
($c_{i,i+1}$ and $c_{i,i+2}$), for comparison. The first column shows that the ratio of monopoles is
almost completely uncorrelated with the other ratios. The ratio of the quadrupoles has a 
significant correlation with the remaining two ratios ($P^{(2)}_1/P^{(0)}_1$ and $P^{(2)}_2/P^{(0)}_2$) 
-- however, the signal-to-noise of the ratio of quadrupoles is insignificant, which means that these
correlations are irrelevant. It should be noted that all these cross-correlations (whether relevant or irrelevant)
were taken into account when we inverted the data Fisher matrix in Section VI.

\bibliography{biblio_alpha_sim}

\end{document}